\documentclass[aps,floats,10pt,superscriptaddress,floatfix]{revtex4}
\usepackage{amssymb,amsmath}
\usepackage{dcolumn}
\usepackage{amsmath,amssymb}
\usepackage{graphicx}
\usepackage{psfrag}
\usepackage{dcolumn}
\usepackage{color}

\def\beq{\begin{equation}}
\def\eeq{\end{equation}}
\def\bea{\begin{eqnarray}}
\def\eea{\end{eqnarray}}
\begin{document}
 \title{Active processes make mixed lipid membranes either flat or crumpled}
\author{Tirthankar Banerjee}\email{tirthankar.banerjee@saha.ac.in}
\affiliation{Condensed Matter Physics Division, Saha Institute of
Nuclear Physics, Calcutta 700064, India}
\author{Abhik Basu}\email{abhik.basu@saha.ac.in,abhik.123@gmail.com}
\affiliation{Condensed Matter Physics Division, Saha Institute of
Nuclear Physics, Calcutta 700064, India}
\affiliation{Max-Planck Institut f\"ur Physik Komplexer Systeme, N\"othnitzer 
Str. 38,
Dresden, D-01187 Germany}
\date{\today}
\begin{abstract}
Whether live cell membranes  show miscibility phase transitions (MPTs), and if 
so, how 
they fluctuate near the transitions remain outstanding unresolved issues
in physics and biology alike.
 Motivated by these questions we construct a generic hydrodynamic theory 
for 
 lipid membranes {{that are active, due for instance, to the 
molecular motors in the surrounding cytoskeleton, or  active protein 
components in the membrane itself}}. We use this to  uncover a direct 
correspondence 
between membrane fluctuations and MPTs. Several testable 
predictions are made: (i)
 generic {\em active stiffening} with 
orientational
long range order (flat membrane) or {\em softening} with crumpling of the 
membrane, controlled 
by the {\em active tension} and (ii) for mixed lipid
membranes, capturing the nature of putative
MPTs 
by measuring the membrane
conformation fluctuations. Possibilities of both first and 
second order MPTs in mixed active membranes are argued for.
Near 
second order MPTs, active stiffening (softening) manifests as a {\em 
super-stiff (super-soft) membrane}. Our 
predictions are 
testable in a variety of {\em in-vitro} systems, e.g., live cytoskeletal 
extracts deposited on liposomes and lipid membranes containing active proteins 
embedded in a passive fluid.

% We discuss plausible experimental
%realizations of our model.
\end{abstract}

\maketitle 
%{\em Introduction:-}
\section{Introduction}\label{intro}

Cell membranes are generally made of several 
lipids and have complex structures~\cite{alberts}. The dynamics 
of cell membranes  are 
affected by biological active (nonequilibrium) processes
(e.g., nonequilibrium 
fluctuations of cell cytoskeletons~\cite{bio-active} and 
active proteins in the lipid 
membrane~\cite{act-prot}); see also Ref.~\cite{john-pump}.
Miscibility phase transitions (MPTs) in equilibrium 
heterogeneous or mixed model lipid bilayers 
and giant
plasma membrane vesicles (GPMVs) are well studied~\cite{rev1,veatch}. In 
contrast, occurrence of MPTs in eukaryotic cell membranes,
remains controversial till
date~\cite{ref}. Whether cellular active processes can control membrane 
fluctuations and associated MPTs in mixed membranes and if so, how, form 
general 
motivations for the present study.

Structural and dynamical complexities of cell membranes preclude 
simple physical understanding of MPTs in cell biological context. This calls 
for studying this question within a simpler nonequilibrium model appropriate 
for an in-vitro setting, where this issue may be addressed systematically and 
possibly verified in suitably designed {\em in-vitro} experiments. To this end,
in this article we construct a hydrodynamic theory for planar 
active 
mixed lipid 
(fluid) membranes~\cite{alberts}. Hydrodynamic approaches have a long 
history of applications in both equilibrium~\cite{chaikin,halpin} and 
nonequilibrium~\cite{sriram-RMP,sriram+john} systems and are successful in predicting 
general physical properties at large scales independent of the 
microscopic (molecular) details of the systems. In particular, our theory is 
applicable to a variety of systems, e.g., a 
lipid bilayer
in an orientable active fluid~\cite{sriram-RMP} in its 
isotropic phase~\cite{iso} or a lipid bilayer with an active component, e.g., 
active proteins, 
immersed in a passive fluid~\cite{act-prot}. We use it to study the membrane
conformation fluctuations and the associated active or nonequilibrium MPTs.  { 
Both second 
order MPT
 (through critical  and tricritical points), and first 
order MPT 
are considered.} We 
uncover a
{\em direct correspondence} between membrane fluctuations and { the nature 
of} the
MPTs, { potentially opening up a new experimental route to study the MPTs}.
  Our predictions are quite 
general; we expect that our characterisations of the membrane fluctuations 
and MPTs should serve as references for  experimental observations on MPTs 
in mixed 
model lipid bilayers and
GPMVs  in isotropic actomyosin extracts with adenosine triphosphate 
(ATP) molecules 
or   solutions of live orientable bacteria~\cite{aran}, and lipid membranes 
with active protein inclusions embedded in passive fluids. From 
perspectives of nonequilibrium physics, our model provides an intriguing 
example where the same underlying microscopic active processes control two 
distinct phenomena, {\em viz.} MPTs of the membrane composition and the nature 
of fluctuations of the membrane conformation, ultimately linking the two in a 
definitive way.

{ In order to focus on the
essential physics of the problem, we consider a planar, tensionless,
two-component,  inversion-symmetric~\cite{inv}, single-layered 
lipid 
membrane~\cite{bilayer} of linear size $L$.}  
In stark contrast to
equilibrium lipid membranes~\cite{luca,tirtha1}, { our model membrane 
displays 
 generic 
stiffening and statistical flatness
for positive active tensions $\sigma_a$, but softening and crumpling 
for $\sigma_a<0$ at any 
temperature $T$}.
{ We describe the planar membrane conformations by a  single-valued 
height field $h({\bf x},t)$  in the Monge 
gauge~\cite{wein,chaikin}, in two dimensions ($2d$) with the  local normal 
${\bf 
n}= 
(-{\boldsymbol\nabla} h,1)=({\delta \bf n},1)$~\cite{wein,chaikin}. 
Then, for positive $\sigma_a$}, 
{away from the critical point for second order MPT 
and across first order MPT} we find variances
\begin{eqnarray}
 \Delta_n&=&\langle (\delta {\bf n}({\bf x},t))^2\rangle\sim const.,\\\Delta_h 
&=& 
\langle h({\bf x},t)^2\rangle\sim\ln L,\label{basicres}
\end{eqnarray}
 in the thermodynamic limit 
(TL). { These imply orientational long range order (LRO), hence statistical 
flatness 
and
positional quasi long range order (QLRO)}; 
here $\langle..\rangle$ implies averaging over noises; 
see dynamical equations~(\ref{heq}) and (\ref{phieq}), respectively, below. 
Near the critical point,
%\end{equation}
the membrane becomes {\em super stiff}: in a mean-field like treatment, we 
show
\begin{equation}
 \Delta_h\sim\ln\ln L,\label{basicres1}
\end{equation}
in TL, an $L$-dependence weaker than in QLRO, which we 
call positional {\em nearly long range order} (NLO). { Moreover, 
$\Delta_n$ 
is  further suppressed near the critical point.}  In contrast, for
 $\sigma_a<0$, $\Delta_n$ and $\Delta_h$ diverge for 
membranes 
larger than a {\em persistence length}, i.e.,  $L>\zeta$~\cite{tirtha1}, 
indicating 
orientational 
and positional short range orders (SRO). {The rest of this article is 
organised as follows. In Sec.~\ref{model}, we set up our coarse-grained 
equations of motion. Then in Sec.~\ref{results} we discuss our results on the 
membrane conformation fluctuations at or across various MPTs. 
Section~\ref{mpts} discusses the various MPTs possible within our model. 
Finally, in Sec.~\ref{summary} we summarise our results. A glossary of our 
results has been added in Sec.~\ref{glossary} for the convenience of the readers. Some 
technical aspects of the calculations involved and a few additional discussions 
related to the main results of this work are made available in Appendices A to 
G for 
interested readers.}

\section{Construction of the model}\label{model}

%{\em Model:-} 
{ We consider an incompressible}
mixed permeable membrane composed of two lipids A and B { of equal amount} 
with
local concentrations $n_A({\bf x},t)$ and $n_B({\bf x},t)$, 
respectively~\cite{tirtha1},
$n_A+n_B=1$.   
The
local inhomogeneity $\phi ~(=n_A-n_B$) is the order parameter for
the MPT. %The local 
%orientation 
%or director fields $\bf p$~\cite{jacques}
%of the embedding isotropic active medium, whose fluctuations
%relax {\em fast}~\cite{iso,fast}, are not hydrodynamic variables.
   Since the active processes may
in general interact differently with A and B, we relax the
usual inversion symmetry of $\phi$ for a binary mixture~\cite{safran} when 
coupled to   local mean curvatures in the
present model (see Refs.~\cite{tirtha1,ayaton,xyz} in this context).

\begin{figure}[htb]
 \includegraphics[width=9cm]{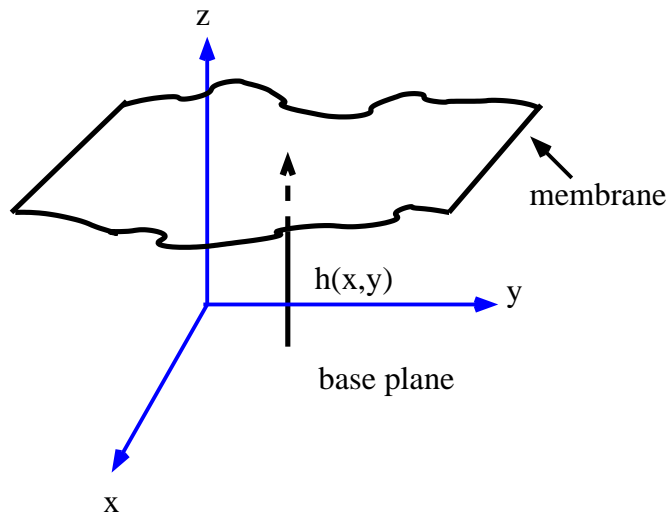}
\caption{Schematic configuration of the nearly flat model membrane in the Monge 
gauge. The membrane is surrounded by bulk fluid above and below.}\label{schem}
\end{figure}

 {Now, consider a nearly flat permeable membrane spread parallel to the 
$xy$-plane, 
i.e., with local normals parallel to the $z$-axis on 
average; see Fig.~\ref{schem} for a schematic diagram.} 
{ While the height 
field $h$ is a nonconserved broken symmetry variable, the order parameter field 
$\phi$ is a conserved density, since $n_A$ and $n_B$ are conserved. The 
relevant equations of motion for $h$ and $\phi$ may be 
derived as follows. The 
membrane, treated as a permeable fluid 
film~\cite{cai-lubensky}, has a local velocity in the normal direction (along 
$z$ direction in this case) $v_{mz}$ given by
\begin{equation}
 v_{mz}=v_{hydro_z} + v_{perm}.
\end{equation}
Here, $ v_{hydro_z}$ is the $z$-component of the local three-dimensional ($3d$)
hydrodynamic velocity ${\bf v}_{hydro}$~\cite{iso}, and $v_{perm}$ represents
the local permeative flows~\cite{iso}.
In general, both $v_{perm}$ and ${\bf v}_{hydro_z}$ may contain equilibrium 
(controlled by a free energy $\mathcal 
F$; see below) and active parts. {The latter contributions cannot be 
obtained from $\mathcal F$. Instead,}
symmetry 
considerations (e.g., translation, in-plane 
rotation, 
inversion symmetry of $h$ and
invariance under tilt for the membrane) may be used to enforce the general 
forms of the
active contribution to $v_{perm}$. We 
write
\begin{equation}
 v_{perm}=\Gamma_h(\lambda\phi^2
+\tilde\lambda\phi)\nabla^2 h -\mu_p \frac{\delta {\mathcal F}}{\delta h} 
\label{vpermact}
\end{equation}
to the lowest order in 
nonlinearities and gradients satisfying the relevant invariances. The first 
term on the rhs of (\ref{vpermact}) with coefficients $\lambda$ and 
$\tilde\lambda$ is the active contribution to $v_{perm}$; $\mu_p$ is a 
(constant)
kinetic coefficient for the equilibrium contribution to the permeative flows. 
The active 
terms with coefficients $\lambda$ and $\tilde\lambda$
are forbidden in 
equilibrium due to 
the
tilt invariance of the associated free energy $\mathcal F$ (see 
below) ~\cite{tirtha1}. They
are, however, permitted here as the tilt invariance in the present
problem must hold at the level of the equations of motion~\cite{sriram+john}. 
All of $\lambda,\tilde\lambda,\mu_p$ vanish for impermeable membranes.
  Furthermore, $\lambda$ and $\tilde\lambda$ control the strength of  
$\sigma_a$, the active tension (see 
below); a non-zero $\tilde\lambda$ models the 
asymmetric dependence of the active processes on lipids A and B. Symmetry 
arguments cannot determine the signs of 
$\lambda,\tilde\lambda$. In this work, we examine the consequences of both 
positive and negative $\lambda$; the sign of $\tilde\lambda$ can be absorbed 
within the definition of $\phi$. 
Further, for a membrane 
with a fixed background  
there are no active contributions to $v_{hydro_z}$ that are more 
relevant than the $\lambda,\tilde\lambda$-terms above~\cite{active-hydro}; 
$v_{hydro_z}=-\Gamma_0\delta{\mathcal F}/\delta h$ with kinetic coefficient 
$\Gamma_0$ being a constant. 
The dynamics of $\phi$ should generally follow a conservation law form 
advection-diffusion equation~\cite{halpin}. To proceed further,  we assume 
$\mathcal F$ for a tensionless, mixed lipid 
membrane to have the { simple} generic form
\begin{eqnarray}
 {\mathcal F} &=& \int d^dx 
[\frac{\kappa}{2}(\nabla^2 h)^2 + \lambda_1 
\phi^2 (\nabla^2 
h)^2 + \lambda_2 \phi (\nabla^2 h)^2\nonumber \\&+& 
   \frac{r}{2}\phi^2 +\frac{1}{2}(\nabla \phi)^2 + \frac{g}{3}\phi^3+ 
\frac{u}{4!} 
\phi^4 
+\frac{v}{6}\phi^6-\tilde h\phi].\label{free-en}
\end{eqnarray}
Here $r=T-T_c$, $u,v>0$. Couplings $g$ and $\tilde h$ can 
be of 
either sign; 
for a symmetric binary mixture $g=0=\tilde h$. Coupling $v$ has 
been added for reasons of thermodynamic stability (see below) and is 
irrelevant in equilibrium with $u>0$. {Free energy 
(\ref{free-en}) without membrane fluctuations ($h=const.$) describes MPTs 
identical to the 
standard liquid-gas phase transition that is generally first order in nature, 
and admits a second order MPT at a critical point that can be accessed only by 
setting $T=T_c$ 
and also tuning $\tilde h$ to a critical value (equivalently setting pressure 
$p=p_c$, the critical 
pressure), in analogy with magnetic systems~\cite{chaikin}. This analogy 
can be made more precise by expanding $\mathcal F$ about $\phi=\phi_0$, with 
$\phi_0$ is 
chosen such that the  $g\phi^3$-term in $\mathcal F$ above vanishes. The 
resulting transformed 
free energy has the form same as that of the Ising model at a finite external 
magnetic field $\tilde h_0$ (related to $\tilde h$ and depends upon the 
chemical potential and temperature) that has generic first order MPTs (or may 
show no transitions) if $\tilde h_0$ is tuned at any general $T$; furthermore, 
second order MPT belonging to the $2d$ 
Ising universality 
class is found if {\em both} $T$ and $\tilde h_0$ are tuned; the corresponding 
critical point is located in a $T-\tilde 
h_0$-plane at $r=0$ or 
$T=T_c$ (in a mean-field description) and $\tilde h_0=0$~\cite{chaikin}.  
In 
fact, the path in the temperature - pressure plane 
of a binary mixture that directly resembles the zero magnetic field path in a 
magnet (which shows a second order transition) is the one with the density 
fixed at the critical density, i.e., the critical isochore. Notice that the 
coexistence curve for $\mathcal F$ above is similar to that for the Ising 
model, except now being asymmetric with respect to the 
order parameter $\langle \phi\rangle =m$ due to the lack of any symmetry of 
$\mathcal F$ under inversion of $\phi$~\cite{chaikin}. The inclusion of 
$h$-fluctuations in 
(\ref{free-en}) via the couplings $\lambda_1,\lambda_2$ does not alter this 
picture~\cite{veatch,tirtha1}. This may be easily seen from the form of 
$\mathcal F$: the $\lambda_1$ and $\lambda_2$ terms effectively only induce 
fluctuation-corrections to $r$ and $\tilde h$, respectively.} To 
what degree this equilibrium physical picture 
is affected by the activity remains to be seen (see below).

 Notice that 
$\mathcal F$ 
implies an 
effective 
composition-dependent bending modulus $\tilde\kappa(\phi)$ of the 
form~\cite{tirtha1}
\begin{equation}
 \tilde\kappa(\phi)= \kappa+2\lambda_1 \phi^2+ 2\lambda_2 \phi.\label{bendphi11}
\end{equation}
Parameters $\lambda_1,\lambda_2$ should be chosen to ensure 
$\tilde\kappa(\phi)>0$ 
for all $\phi$, that guarantees a thermodynamically stable flat phase, 
{given by the minimum of $\mathcal F$,} in 
equilibrium. We choose $\lambda_1>0$~\cite{tirtha1}. The sign of $\lambda_2$ is 
arbitrary and can be absorbed within the definition of $\phi$; for 
concreteness we choose $\lambda_2>0$. While the individual signs of 
$\tilde\lambda$ and $\lambda_2$ are arbitrary, the sign 
of the product $\lambda_p\equiv \tilde\lambda\lambda_2$ is crucial { to 
what follows below} and controls the ensuing
macroscopic behaviour. 

Putting together everything, the
dynamical equations for $h$ (with $\partial_t h = v_{mz}$) and $\phi$  to the 
lowest order in spatial 
gradient
expansions, in the long wavelength limit take the forms (for a fixed background 
medium)
\begin{eqnarray}
 \frac{\partial h}{\partial t}&=&\Gamma_h [-\kappa\nabla^4 h +
(\lambda\phi^2 + \tilde \lambda\phi) \nabla^2 h ]+ f_h,\label{heq}\\
\frac{\partial\phi}{\partial t}&=&\Gamma_\phi 
\nabla^2[r\phi -\nabla^2\phi +
\frac{u}{3!} \phi^3 +2\lambda_1 \phi(\nabla^2 h)^2 \nonumber \\&+& \lambda_2
(\nabla^2 h)^2+v\phi^5] + {\boldsymbol\nabla}\cdot{\bf f}_\phi.\label{phieq}
\end{eqnarray}
Notice that  the 
terms with coefficients $\lambda_1,\lambda_2$ in (\ref{free-en}) generate 
additional equilibrium terms in (\ref{heq}); these are however
subleading in the hydrodynamic limit (in a scaling sense) to the active
$\lambda,\tilde\lambda$-terms, respectively, and are hence omitted from 
(\ref{heq})~\cite{omit}.  Kinetic 
coefficient $\Gamma_h=\mu_p+\Gamma_0$ is a constant for a membrane 
with a fixed background; 
$\Gamma_\phi$ is also a constant.  Notice that separately Eq.~(\ref{phieq}) may 
be 
wholly obtained from $\mathcal F$ and is just of the 
``model 
B'' 
conservation law form (in the nomenclature of Ref.~\cite{halpin}); this is 
because there are no  active terms which are more relevant (in a scaling 
sense) than those already
included in (\ref{phieq}). {{ In fact, the $\lambda_1-$ and 
$\lambda_2-$ 
nonlinear terms in (\ref{phieq}) originate from the composition-dependent 
bending modulus $\kappa(\phi)$ in (\ref{bendphi11}) in the free energy 
(\ref{free-en}).}}  {This does not, however, in general imply that $\phi$ 
follows an equilibrium dynamics; its coupling with $h$ ensures that the 
resulting effective dynamics for $\phi$ is detailed balance breaking.} 
Further, we have 
ignored any in-plane advection of $\phi$ for simplicity~\cite{active-phi}. 
Noises $f_h$ 
and 
${\bf
f}_\phi$ are zero-mean, Gaussian distributed with variances given by
\begin{eqnarray}
\langle 
f_h 
({\bf q},\omega) f_h ({\bf
q'},\omega')\rangle &=& 2D_h\Gamma_h \delta ({\bf q+q'})\delta
(\omega+\omega'),\\
\langle f_{\phi i}({\bf q},\omega) f_{\phi
j}({\bf q'},\omega')\rangle &=& 2D_\phi\Gamma_\phi\delta_{ij}\delta
({\bf q+q'})\delta (\omega+\omega').
\end{eqnarray}
Here, $D_h\neq D_\phi$ in general; ${\bf 
q,q}'$ are wavevectors and $\omega,\omega'$ are frequencies, $q=|\bf q|$.
 Noises $f_h$ and $f_{\phi i}$ should contain both thermal as well as 
active contributions.

  %The form
%of (\ref{phieq}) produces standard SMPT belonging to the 2d Ising
%universality class in the equilibrium limit
%($\lambda=0=\tilde\lambda$)~\cite{veatch}. %: clearly, 
%(\ref{heq})
%and (\ref{phieq}) are %invariant
%under a tilt $h\rightarrow h+ {\bf a}\cdot {\bf x}$, where $\bf a$ is any
%constant vector.

\subsection{Active terms}\label{active}

The dynamical equations (\ref{heq}) and (\ref{phieq}) are 
constructed using symmetry arguments. As a result, these serve as good 
hydrodynamic representations for a variety of systems that conform to the same 
symmetries as Eqs.~(\ref{heq}) and (\ref{phieq}). Equivalently,
the active terms in 
(\ref{heq}) can be motivated in various physical contexts.
{For instance}, consider an inversion-symmetric, mixed, planar
fluid membrane placed in an isotropic, active suspension of actin
filaments~\cite{sriram-RMP}, grafted normally to it.  This is imposed by 
the condition
%Now assume that the filaments are grafted normally on the membrane
 ${\bf p}\cdot {\bf  n}=1 $, where $\bf p$ is the local 
orientation 
or director fields~\cite{jacques} which describe the local orientation of the 
actin filaments. 
This yields
$p_j=\partial_j h \,(j=x,y)$ to the linear order in height
fluctuations~\cite{epje3} {at the location of the membrane ($z=h$)}.  
In the  embedding bulk isotropic active medium, there is no net 
orientational order and hence the fluctuations of $\bf p$
relax {\em fast}~\cite{iso,fast}. Thus, $\bf p$ in the bulk are not 
hydrodynamic variables and can be ignored in the long time limit as far as the 
bulk embedding fluid is concerned. At $z=h$, the location of the membrane, 
however, $\bf  p$ is nonzero and is slaved to the membrane fluctuations, as 
above. 
The
general form of  the $z$-component of the local  
membrane velocity, taking into account the permeative flow,
is 
\begin{eqnarray}
\frac{\partial h}{\partial t}&=&v_z(z=h)=-\mu_p\delta {\mathcal F}/\delta h 
+ 
X(\phi)\partial_j(p_z p_j) + v_{hydro_z}\nonumber \\
&=&-\Gamma_h\delta F/\delta h  + X \nabla^2 h
\end{eqnarray}
at the lowest order in 
fluctuations (see Ref.~\cite{iso} for a similar 
active contribution), consistent with the 
inversion-symmetry.  Now, with  
$X(\phi)=\Gamma_h(\lambda\phi^2
+\tilde\lambda\phi)$~\cite{alpha} and $p_z=1$ to the leading order in 
smallness, we 
recover 
(\ref{heq}). {{ That this local orientation 
fluctuation gives rise to active 
permeation flows is not surprising:   
the
polymerisation/depolymerisation and treadmilling of the actin 
filaments, which are active processes~\cite{iso,sykes,alberts}, pull or push 
the membrane. This contributes to the permeation flow and can either 
reinforce
or oppose the corresponding equilibrium contribution.}
%However, there is no bulk 
%propulsion for a symmetric membrane. 
{{Yet another system that may be described by (\ref{heq}) and 
(\ref{phieq}) is the hydrodynamics of permeable lipid membranes with 
active protein inclusions~\cite{act-prot,gov22} that are either embedded in the 
bilayer or
adsorbed to it (e.g., cytoskeleton proteins) and can phase separate, 
immersed in a passive fluid. These active
proteins convert chemical energy of the ATP molecules or of the light 
shone~\cite{BR-protein} on the membrane into mechanical motion 
of the membrane.  Local order parameter
$\phi$ in such  systems should describe the mole fraction of the active 
components. The main physical features of these active proteins are
that they force the membrane locally and independently
of each other, generating a local normal motion of the membrane that should 
evidently depend upon $\phi$~\cite{gov22}. The active terms in (\ref{heq}) then 
simply 
model the $\phi$-dependence of the local normal velocity of the membrane. This 
$\phi$-dependence leads to the active tension $\sigma_a$; see below.}}

Now imagine regions of 
nonzero mean curvature with excess lipid of one kind so that $\phi$ picks up a 
non-zero value with a specific sign. Such a region then 
either pulls up the curved region further (instability) or tries to flatten 
the curvature (stable membrane) due to the active processes; see 
Fig.~\ref{modellam} for a schematic picture. 
\begin{figure}[htb]
\includegraphics[width=9cm]{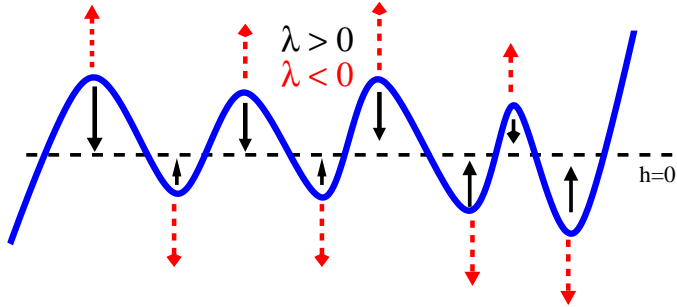}
\caption{(Color online) Schematic representations of the active  
velocity
($\tilde\lambda=0$) of the 
membrane (blue curved line) for $\lambda>0$ (stabilising, solid black arrows) 
and $\lambda<0$ 
(destablising, broken red arrows). A non-zero $\tilde\lambda$ leads to
further $\phi$-dependent modification of the active velocity (not shown). }  
\label{modellam}
\end{figure}
 For an 
active fluid with actin filaments, $\lambda,\tilde\lambda$ 
should 
scale with the concentration $C_0$ of the ATP molecules 
and 
 the free energy released in hydrolysis of 
ATP, $\Delta\mu\sim 0.8\times10^{-18}\,kJ$~\cite{epje4}.  In the example of 
a live cytoskeletal extract, active contributions to $v_{perm}$ should depend 
on the 
treadmilling speed of the actin filaments$~\sim O(1\; \mu 
m/h)$~\cite{tread}; this may be used to make an estimate 
on 
$\lambda,\tilde\lambda$. { For a membrane with an active component, 
$\lambda,\tilde\lambda$ should scale with the mean concentration of the active 
species in the membrane.}

It is now instructive to compare and contrast with the generic leading 
order active terms present for a pure tensionless membrane ($\phi=0$).
For a pure membrane, the leading order active propulsion velocity  is just a 
constant: $X(\phi=0)=\Gamma_h\alpha$ to the leading order. 
Thus,
for such a pure membrane with a vanishing tension in an active medium, 
the dynamical equation for $h$
 can be obtained  
from an 
effective free energy that now includes an {\em effective surface tension} 
$\sigma'=\alpha$ in the 
free energy. We can write
\begin{equation}
 \frac{\partial h}{\partial t}=-\Gamma_h \frac{\delta F_h}{\delta h} 
+v_{hydro_z}+ f_h,
\end{equation}
where $F_h = \int d^2x [\sigma'({\boldsymbol\nabla} h)^2/2 + \kappa 
(\nabla^2 h)^2/2]$.
Thus, for a symmetric pure membrane, the active effects may be wholly described 
by a 
modified equilibrium 
free energy $F_h$ to the leading order, or, equivalently, the role of 
active effects here is to just to introduce an effective surface tension. 
For a mixed 
membrane, there is no such 
general equivalence 
with a simply
modified equilibrium model.

%At equilibrium $D_h=D_\phi=T$, the thermody temperature.
%However, the present model being out of equilibrium, $D_h=D_\phi$ in general
%and have no physical interpretation of $T$.

%{\em Results:-} 
\section{Results}\label{results}

{{In this Sec. we first derive our results on membrane 
fluctuations without any (equilibrium) surface tension by using the model 
equation (\ref{heq}). We analyse Eq.~(\ref{heq}) in a mean field-like spirit 
and compare its properties with an isolated tensionless fluid membrane at 
thermal equilibrium.  We then establish their correspondence with the order of 
MPTs, the principal prediction of this work. Next, we briefly touch upon how a 
finite surface tension may affect our results. We now proceed to discuss these 
in details below.}}

\subsection{Properties of membrane fluctuations}

The lack of knowledge about 
the order 
of 
MPTs in a symmetric mixed membrane embedded in an active 
fluid, demands that we must allow for the 
possibility of both first order MPT  and second order MPT, and study their 
connections with the 
membrane fluctuations separately; { see Ref.~\cite{recent} for a recent 
study of first order MPT in a lipid bilayer in presence of transmembrane 
proteins}. {
In particular, this opens the intriguing possibility of activity-induced first 
order MPT 
in a mixed lipid bilayer that admits only second order MPT in the equilibrium 
limit.} We 
present theoretical arguments in favour of 
both first order MPTs and second order MPTs later in the text.  
From (\ref{heq}), we extract  an {\em active tension} $\sigma_a$. {\ We 
write in the Fourier space
\begin{equation}
 \frac{\partial h}{\partial t} = \Gamma_h [-\kappa q^4 h - 
X(\phi) q^2 h] + f_h,\label{newh}
\end{equation}
where}
{  $X(\phi) = \lambda\phi^2 + \tilde\lambda\phi$; ${\bf q}$ is a Fourier 
wavevector. Now write $X(\phi)=\langle 
X(\phi)\rangle +\delta X(\phi)$; $\delta X(\phi)$ is the fluctuation of 
$X(\phi)$ about its mean $\langle X(\phi)\rangle$. Then, 
neglecting $\delta 
X(\phi)$ in comparison with $\langle X(\phi)\rangle$ for (assumed) small 
fluctuations} (in mean-field like treatment) and
 for any $\lambda\neq 0$, { we can extract an {\em active tension} 
$\sigma_a$ as follows:} 
\begin{eqnarray}
\;
%\label{kappaeff}\\
\sigma_a = \lambda\langle\phi^2\rangle,\label{activeten}
\end{eqnarray}
with $\langle \phi\rangle =0$ for the whole system with equal 
A and B at all $T$; clearly
$\sigma_a$ is positive (negative) for $\lambda > 
(<)0$. {Consider $\lambda >0$ first.} { Equations (\ref{newh}) and 
(\ref{activeten}) imply for 
the membrane height fluctuations
\begin{equation}
 C_h(q)=\langle |h({\bf q})|^2\rangle =\frac{D_h}{\sigma_a q^2 +\kappa 
q^4}.\label{hcorr} 
\end{equation}
{{Thus $C_h(q)$ is dominated by the active tension $\sigma_a 
q^2>0$ in the long wavelength limit, since $\kappa q^4$ is subleading to it (in 
a scaling sense). At this stage it is formally instructive to compare $C_h(q)$ 
as given 
in (\ref{hcorr}) 
 with that of a tensionless fluid membrane (since our model membrane has 
zero surface tension in equilibrium) with an {\em 
effective bending modulus} $\kappa_e$ in equilibrium at temperature 
$D_h/(\Gamma_h K_B)$. This
yields}}
\begin{equation}
 \kappa_e(q)=\kappa +\frac{\sigma_a}{q^2}=\kappa+\frac{\lambda}{q^2}\langle 
\phi^2\rangle.\label{kappaeff}
\end{equation}
{{ Unsurprisingly, $\kappa_e(q)$ has a part that diverges as 
$1/q^2$, a reflection of the active tension $\sigma_a(q)$, that is 
dimensionally identical to the usual surface tension. Equivalently, to the 
leading order our tensionless model active membrane behaves like an 
equilibrium 
membrane under tension. }}
This suggests generic stiffening (softening) of the model membrane
for $\lambda > (<) 0$ at {\em any} $T$, in contrast to an isolated fluid 
membrane in equilibrium~\cite{tirtha1}. Positive and negative 
$\lambda$, respectively, physically imply that creation of nonuniform regions with 
specific signs of $\phi$ (i.e., A- or B-rich domains) should make the membrane 
either try to flatten out ($\sigma_a >0$), or curve more ($\sigma_a <0$); 
see also Refs.~\cite{iso,john-pump} for active tension in different models for 
active membranes.
Now 
assume $\lambda>0$. In the ordered phase, 
$\langle\phi^2\rangle=m^2$ (neglecting fluctuations) is larger than its value in the 
disordered phase; hence  
$\kappa_e(q) >\kappa$ in
the ordered phase, where $m$ is the average of $\phi$ in an A- or B-rich 
domain in the ordered phase.
%Despite the similarity in the generic
%behavior of $\kappa_e(q)$ in the disordered and ordered phases as outlined
%above, interestingly there are significant differences in the forms of
%$\kappa_e(q)$ as $T_c$ ($T^*)$ is approached the transition temperature is
%approached, assuming SMPT (FMPT).

For a putative first order MPT at $T=T^*$, we write 
\begin{eqnarray}
&&\sigma_a(q)=\lambda\langle\phi^2\rangle\implies\kappa_e (q)=\kappa 
+\lambda 
\langle\phi^2\rangle/q^2~(T>T^*),  \nonumber \\
&&\sigma_a(q)=\lambda m^2\implies\kappa_e(q) = \kappa + \lambda
m^2/q^2~ (T < T^*),
\end{eqnarray}
ignoring 
$\phi$-fluctuations in comparison with $m^2$ for $T < T^*$. Thus 
there is a {\em jump} in
$\kappa_e$, that is large for small $q$, as $T$ crosses $T^*$. Now,
%\begin{eqnarray}
%\Delta_n&=&\int_{1/L}\frac{d^2
%q}{(2\pi)^2}\frac{D_h}{
%\kappa_e (q)q^2},\,
%\Delta_h= \int_{1/L}\frac{d^2
%q}{(2\pi)^2}\frac{D_h}{
% \kappa_e(q)q^4}.\nonumber
%\end{eqnarray}
%Thus,
\begin{equation}
 \Delta_h = D_h \int_{2\pi/L}^{\Lambda} \frac{d^2 q_1}{\kappa_e q_1^4} \approx 
\tilde A\ln L
\end{equation}
in TL, implying positional QLRO, { where, $\tilde A$  is a nonuniversal 
constant 
with a value dependent upon ordered and disordered phases; 
$\tilde A = 
\frac{D_h}{2\pi \lambda 
m^2}~({\rm for}\,T<T^*)$; see Eq.~(\ref{basicres})}. Here, ${\boldsymbol q_1}$ 
is a wavevector;
$|{\boldsymbol q_1}|=q_1$; $\Lambda$ is an upper wavevector cut-off.
With  Eq.~(\ref{activeten}), 
\begin{equation}
\Delta_n = D_h \int^{\Lambda}_{2\pi/L}\frac{q_1^2 
d^2 
q_1}{(2\pi)^2\kappa_e q_1^4}
\end{equation}
is finite in TL (i.e., 
orientational LRO), independent of the nature of MPTs. 
 Note that both $\Delta_n,\,\Delta_h$ are 
{\em discontinuous} across $T^*$, due to the discontinuity in $\kappa_e(q)$ 
across first order MPT.

In contrast, for second order MPT
 $\langle \phi^2\rangle$ changes continuously on both sides of
$T=T_c$ and rises as $|r\,(T)|$ becomes smaller. Thus
$\kappa_e(q)$, as given by  (\ref{activeten}),  rises 
smoothly as $T_c$ is approached from either side; see Fig.~\ref{kappa}.

%\widetext
\begin{widetext}
\begin{figure}[h]
\includegraphics[width=16cm]{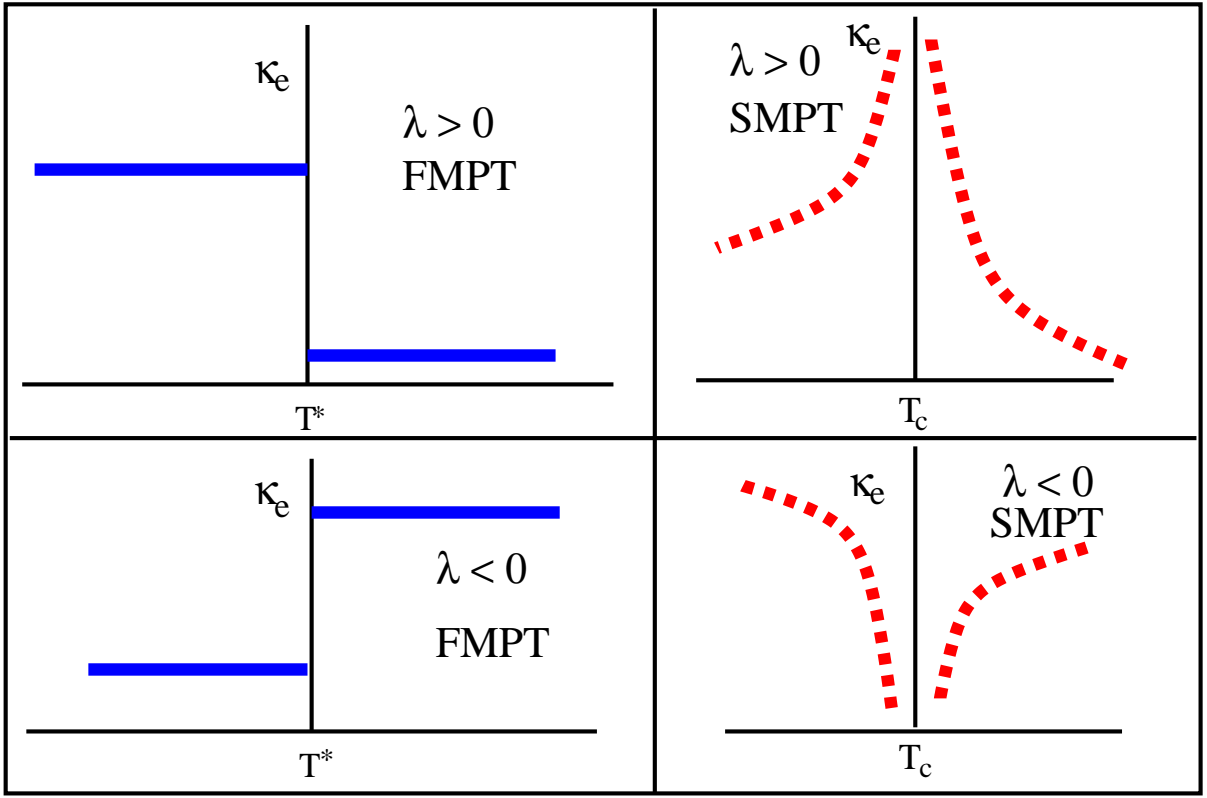}
\widetext{\caption {(Color online) Schematic variation of $\kappa_e 
(q=q_0)$ 
across first order MPT (left) and second order MPT (right) for both $\lambda>0 
(\sigma_a>0)$ 
(top) and $\lambda<0 (\sigma_a<0)$ 
(bottom) with  $q_0= 2\pi/L$; membrane size $L$ is chosen such that $\kappa_e 
(q_0)>0$ with $\lambda<0$, or $L<\zeta$, the persistence length.  Smooth 
(discontinuous) variations of $\kappa_e$ across $T_c$ ($T^*$) 
are shown (see text).}}  \label{kappa}
\end{figure}
\end{widetext}

%\narrowtext

%\newpage

For $T > T_c$, 
\begin{equation}
\kappa_e(q) \approx \frac{\lambda D_{\phi}}{4\pi^2 
q^2}\int^{\Lambda}_{2\pi/L}\frac{d^2q_1}{r+q_1^2} =\frac{\lambda D_{\phi}}{4\pi 
q^2} \ln |\frac{r+\Lambda^2}{r}|.
\end{equation}
This yields
\begin{equation}
 \Delta_h=\tilde G\ln L
\end{equation} 
 in TL, implying positional 
QLRO for $T > T_c$; giving Eq.
(\ref{basicres}) above; $\tilde G= \frac{2 D_h}{\lambda D_\phi \ln |\frac{r+\Lambda^2}{r}|}$, a 
nonuniversal constant, different from $\tilde A$. For 
$T<T_c$ as well, positional QLRO 
holds, however, with a different nonuniversal value for $\tilde G$.
For both $T > T_c$ and $T < T_c$,
\begin{equation}
\Delta_n=\int^{\Lambda}_{2\pi/L}\frac{ d^2q}{(2\pi)^2}\frac{D_h}{\kappa_e 
q^2}
\end{equation}
remains finite in TL.
However, unlike for first order MPT, both $\Delta_n$ and $\Delta_h$ are 
continuous across $T_c$ 
for second order MPT.  Variations of
$\kappa_e(q)$ around first order MPT and second order MPT for a given $q$ are 
shown schematically in
Fig.~\ref{kappa}; also see Glossary at the end summarising our 
results on 
$\kappa_e(q),\,\Delta_n,\,\Delta_h$ across MPTs.

 Care must be taken while analysing the membrane fluctuations close to 
the critical point: large  $\phi$-fluctuations very close to the critical point 
should qualitatively change 
 $\sigma_a(q)$ and hence
$\kappa_e(q)$. For 
simplicity, set $\tilde\lambda=0=\lambda_2$, such that the 
Ising
symmetry for $\phi$ is restored. This 
suffices for our purposes here, since we are interested in second order MPT 
only.  
From (\ref{activeten}) with $\lambda>0$ and within a linearised 
approximation to Eq.~(\ref{phieq}), we 
find near the critical point ($r\approx 0$)
\begin{eqnarray}
\sigma_a(q)&=&-\frac{\lambda D_\phi}{2\pi }\ln q\nonumber \\
\implies \kappa_e(q)&=&\kappa  -\frac{\lambda D_\phi}{2\pi q^2}{\rm ln} 
q \approx  -\frac{\lambda D_\phi}{2\pi q^2}{\rm ln} q,\label{effkappa1}
\end{eqnarray}
for small $q$; { see Appendix
for a renormalised version of Eq.~(\ref{effkappa1}).}
 Then, 
 \begin{equation}
 \Delta_n=
 -\frac{D_h}{\lambda D_\phi} \int^{\Lambda}_{2\pi/L} \frac{q dq}{{\rm ln}~q}\sim 
{\rm finite}
\end{equation}
and
\begin{equation}
\Delta_h=(D_h/\lambda D_\phi)\ln\ln L
\end{equation}
in TL. This
%\end{equation}
 establishes orientational LRO and positional NLO, respectively, with 
$D_h/(\lambda D_\phi)$ as the 
nonuniversal amplitude; see
Eq.~(\ref{basicres1}). 
Thus at $T_c$, $\Delta_n$ is further suppressed { from its value 
at $T\neq T_c$}, yielding a {\em super-stiff} membrane
at the critical point.  This result
holds with or without the ambient fluid hydrodynamics.  
 A schematic phase diagram in the $\lambda-T$ plane is shown in 
Fig.~\ref{lamT}. { These are in stark contrast with their equilibrium 
results.} In 
equilibrium 
$\Delta_n$ scales as $\ln\ln L$ at
$T=T_c$~\cite{tirtha1}, displaying orientational NLO; at all other $T$, 
a pure
membrane in equilibrium
does not remain flat at large scales~\cite{tirtha1}.

For $\lambda<0$, we have $\sigma_a<0$. Hence $\kappa_e(q)<0$ for sufficiently 
low 
$q$, implying long
wavelength instability for planar membranes or occurrence of {\em membrane 
crumpling}. 
In general, larger $\langle\phi^2\rangle$  in the 
ordered phase leads to a smaller $\kappa_e(q)$. We
define a persistence length $\zeta$, such that for $q=2\pi/\zeta$,
$\kappa_e(\zeta)=0$~\cite{tirtha1,o-n}. { This clearly indicates instability 
of flat 
membranes, and as argued in Ref.~\cite{luca}, the membrane gets crumpled. }
Physically, for length 
scales  $L<\zeta$ the
membrane appears flat on average, where as for 
$L>\zeta$, it is crumpled. { The strong dependence of $\kappa_e$ on the
nature of the transition (first order MPT or second order MPT) is also reflected 
in $\zeta$.}
In particular, across an first order MPT at $T^*$,
\begin{eqnarray}
 &&\zeta=2\pi\sqrt{\frac{\kappa}{|\lambda|\langle\phi^2\rangle}},\;\; {\rm for} 
\;T>T^*\; {\rm and} \nonumber \\&&  \zeta= 
2\pi\sqrt{\frac{\kappa }{ |\lambda|m^2}} \;\; {\rm for}\; T<T^*,
\end{eqnarray}
thus showing a jump in $\zeta$; in general $\zeta (T>T^*) > \zeta (T<T^*)$. In 
contrast, there is no discontinuity in $\zeta$ for
second order MPT at $T=T_c$: $\zeta$ satisfies 
\begin{equation}
\zeta^2 \ln [\zeta/(2\pi)] = \frac{8\pi^3\kappa}{D_{\phi}|\lambda|} 
\label{zetaeq}
\end{equation}
at $T_c$.  Away from $T=T_c$, $\zeta (T>T_c)>\zeta( T<T_c)$, similar to the 
behaviour of $\zeta$ across $T=T^*$ in first order MPT.
%Variation of $\zeta$ with $T$ 
%across
%first order MPT and SMPT are shown schematically in Fig.~\ref{zeta}. 
%\begin{figure}[htb]
% \includegraphics[width=7cm]{model.eps}
%\caption{(Color online) Schematic phase diagram in the $\lambda-T$ plane. 
%Super stiff and super crumpling lines are marked.}  \label{zeta}
%\end{figure}
Both $\Delta_n$ and $\Delta_h$ diverge at finite $L\sim\zeta$, 
implying orientational and positional SRO. Due to the large 
 fluctuations of $\phi$ at the critical point, $\zeta (T_c)\ll \zeta (T\neq 
T_c)$, giving {\em super-crumpling} of the membrane, in contrast to super 
stiffness for $\lambda>0$ at the critical point; see {Appendix for more 
details}.

\subsection{ Correspondence
between membrane fluctuations and order of MPTs - experimental implications}

Consider now the implications of the above results on the 
measurements of membrane conformation fluctuations. These may be measured by 
standard spectroscopic methods, see, e.g., Ref.~\cite{timo}. Notice that the 
knowledge of the behaviour of $\sigma_a=\lambda\langle\phi^2\rangle$, or 
$\kappa_e(q)\equiv \kappa+\lambda\langle 
\phi^2\rangle/q^2$ immediately enables us to find the scaling of $C_h(q) 
=D_h/(\sigma_a 
q^2 + \kappa q^4)$ 
across second order MPT or first order MPT, that can be measured in experiments. 
We make the following 
general conclusions:

(i) With 
second order MPT at both $T>T_c$ or $T<T_c$, $C_h(q)\sim 1/q^4$ for large $q$, 
where as 
$C_h(q)\sim 1/q^2$ for small $q$ and $\lambda>0$; $C_h(q)$ diverges for 
$q\rightarrow 0$ only with no finite wavevector singularities. Further, since 
$\sigma_a 
(T<T_c)>\sigma_a(T>T_c)$, $C_h(q)(T<T_c)<C_h(q)(T>T_c)$ for sufficiently small 
$q$, when $\sigma_a q^2$ dominates over $\kappa q^4$. Furthermore, the 
difference $C_h(q)(T>T_c) - C_h (q) 
(T<T_c)$ vanishes as $T\rightarrow T_c$, i.e., $C_h(q)$ has no 
discontinuity as a function of $T$, in agreement with the continuity of 
$\sigma_a$ or $\kappa_e$ across $T_c$. 

(ii) In contrast, for $\lambda <0$ and with second order MPT,  
$C_h(q)(T<T_c)> C_h(q)(T>T_c)$. In addition,
$C_h(q)$ diverges at a finite wavevector $q_c\sim 2\pi/\zeta$. Nonetheless, 
$C_h(q)$ remains continuous across $T=T_c$ even with $\lambda <0$. 
These results are summarised in the form of schematic figures in Fig.~\ref{ch}.

(iii) In case of first order MPT, $C_h(q)(T<T^*)<C_h(q)(T>T^*)$ with 
$\lambda>0$, and 
$C_h(q)(T<T^*)>C_h(q)(T>T^*)$ with $\lambda <0$. However, unlike across second 
order MPT, 
the difference $C_h(q)(T>T^*) - C_h (q) (T<T^*)$ {\em does not} vanish as 
$T\rightarrow T^*$. Thus, $C_h(q)$ is {\em discontinuous} across $T=T^*$, a 
consequence of the discontinuity of $\sigma_a$ or $\kappa_e$. Qualitatively, the behaviors
across the transition temperature $T^*$ are similar to those for second order MPTs.

\begin{figure}[htb]
 \includegraphics[width=9cm]{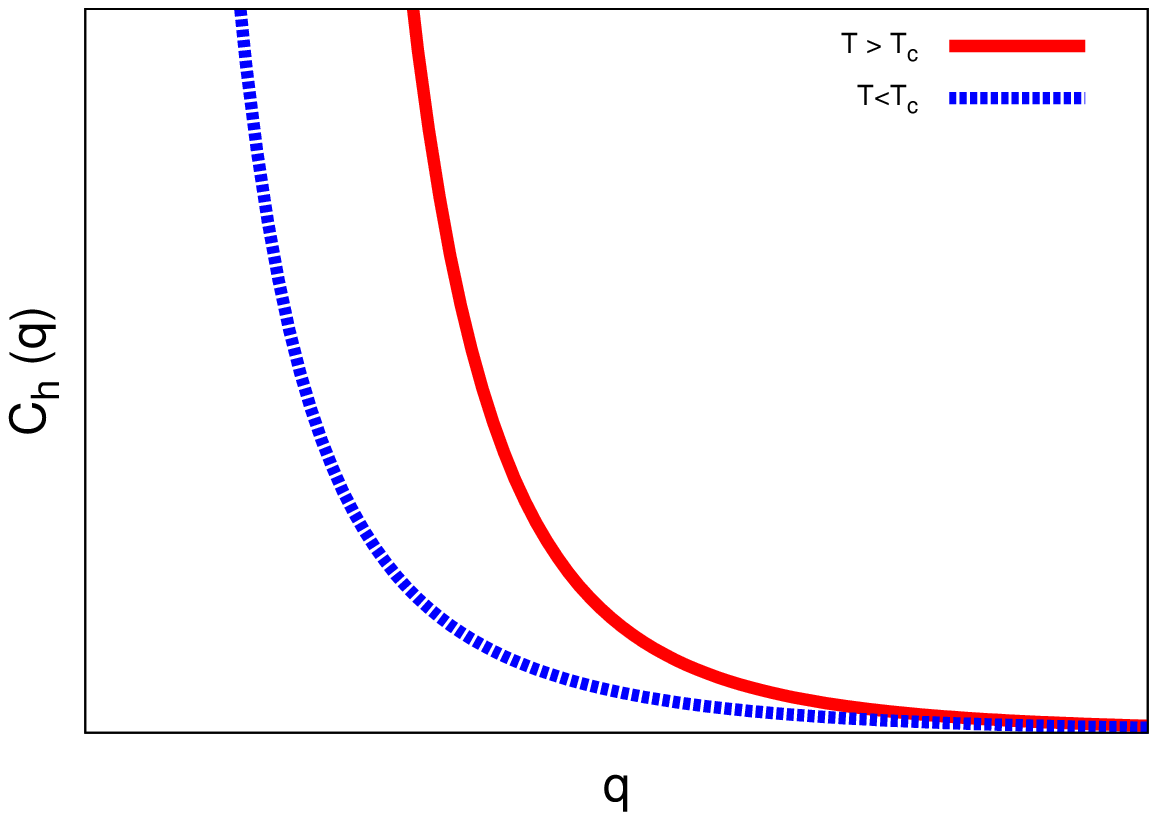}\\
 \includegraphics[width=9cm]{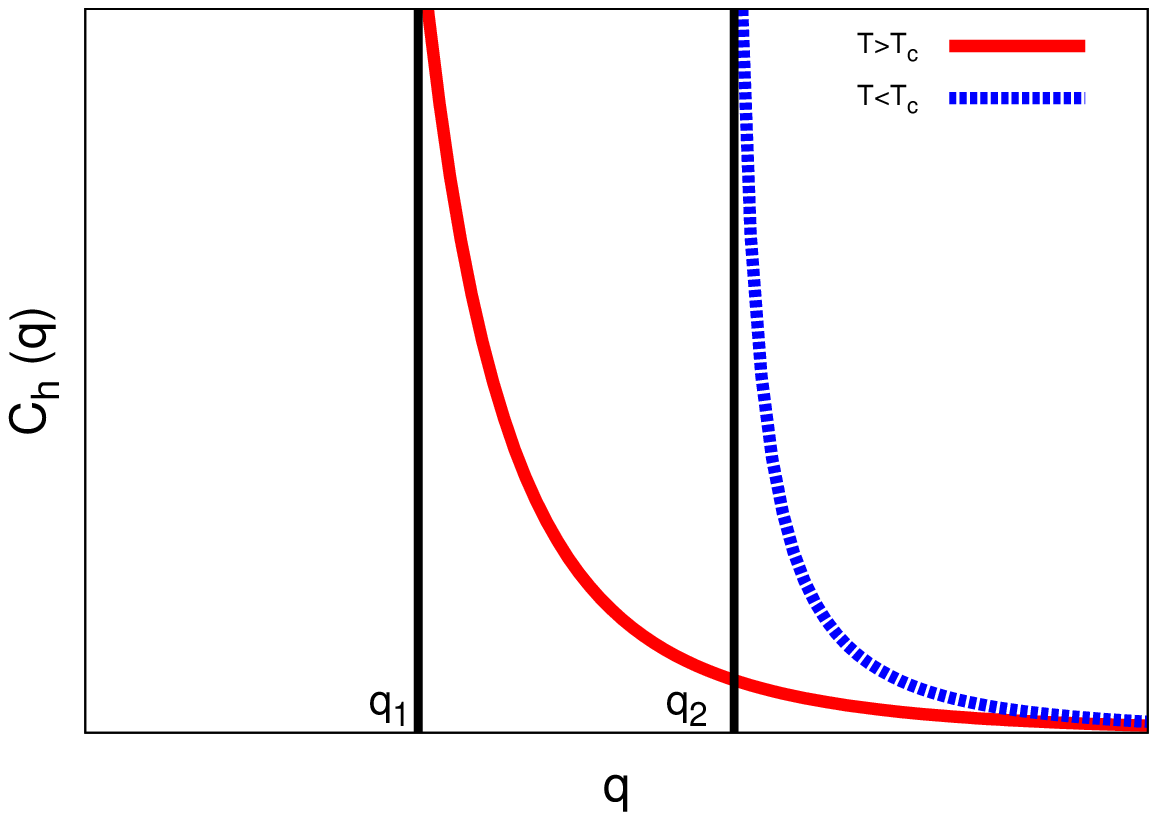}
 \caption{Schematic scaling of $C_h(q)$ with $q$ across second order MPTs: 
(top) $\lambda>0$: 
 red (solid) and blue (broken) curves represents $C_h(q)$ in the disordered 
($T>T_c)$ and ordered  ($T<T_c$) phases, respectively; $C_h(q)$ diverges 
only when $q\rightarrow 0$ for both $T>T_c$ and $T<T_c$ (see text). (bottom) 
$\lambda<0$:
 red (solid) and blue (broken) curves represents $C_h(q)$ in the disordered 
($T>T_c)$ and ordered  ($T<T_c$) phases, respectively. Finite wavevector 
singularities of $C_h(q)$ at wavevectors
  $q_1=2\pi/\zeta_{T > T_c}$ and $q_2=2\pi/\zeta_{T < T_c}$ with 
$\zeta$ 
following Eq.~(\ref{zetaeq}) are visible. }\label{ch}
\end{figure}

\begin{figure}[htb]
\includegraphics[width=9cm]{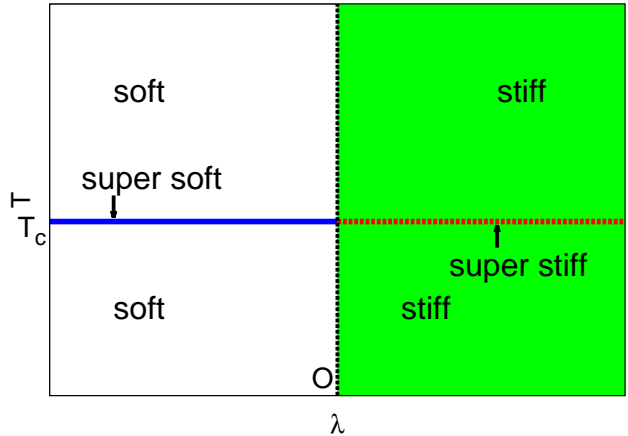}
\caption{(Color online) Schematic phase 
diagram in the $\lambda-T$ plane ($\tilde\lambda=0$). 
Super-stiff and super-crumpling lines are marked. The solid horizontal  
(blue) and broken vertical lines
refer to $T=T_c$ and $\lambda=0$, respectively. Symbol $O$ marks the origin $(0,0)$.}  \label{lamT}
\end{figure}
%{\em SOMPT and membrane stiffening:-}  Set $\lambda>0$.
 { The changes in the membrane 
conformations with $\lambda$, characterised by $\Delta_n,\,\Delta_h$ at a fixed 
$T$ may 
be viewed as 
 a {\em nonequilibrium structural phase transition}} between soft and stiff phases 
{ (see 
Fig.~\ref{lamT})}. An 
order parameter 
for this transition may be constructed as in~\cite{tirtha1,o-n}.

\subsection{Effects of a finite surface tension}\label{tension}

{If the membrane has a finite surface tension $\sigma >0$, then 
it generically suppresses $h$-fluctuations. 
Equation~(\ref{heq}), in the presence of a finite surface tension $\sigma$, now 
modifies to 
\begin{equation}
 \frac{\partial h}{\partial t}=\Gamma_h [-\kappa\nabla^4 h +
(\sigma + \lambda\phi^2 + \tilde \lambda\phi) \nabla^2 h ]+ f_h,\label{newheq}
\end{equation}
The equilibrium terms (\ref{newheq}) can obtained from the free energy 
${\mathcal 
F}$, {now supplemented by} contributions from $\sigma$.
Extracting an active tension $\sigma_a$ from (\ref{newheq}) using the 
logic outlined above, we find that the dynamics is 
controlled by
the total tension $\sigma_{tot}$ that includes both the active 
tension and $\sigma$, 
\begin{equation}
\sigma_{tot}=\sigma + \sigma_a > (<) \sigma\;\;{\rm for}\;\; 
\sigma_a > (<) 0, 
\end{equation}
where $\sigma_a$ is defined as in Eq.~\ref{activeten}. 
A positive $\sigma_{tot}$ ($\lambda>0$) necessarily suppresses membrane 
fluctuations. Thus, 
for $\sigma_a>0$, the role of a non-zero $\sigma$ is to suppress membrane 
fluctuations further. For $\sigma_a <0$ ($\lambda<0$), crumpling instabilities 
should set in only
for $\sigma_{tot}=0$. Ignoring $\tilde\lambda$, this now yields a non-zero 
threshold of instability for $|\lambda |\propto \sigma$ for all $T$ with first 
order MPT 
and $T\neq T_c$ with second order MPT:
\begin{equation}
 \lambda = - \frac{\sigma}{\langle\phi^2\rangle}
\end{equation}
 determines the instability threshold of $\sigma$.
For $T\approx T_c$ with second order MPT, $\langle\phi^2\rangle$, and hence 
$\sigma_a$ 
diverges 
with $L$ as $\ln L$; thus the threshold of instability of $\lambda$ vanishes in 
TL. In the more general case, one may consider a composition-dependent 
surface tension $\sigma(\phi)$. Using a simple form for $\sigma(\phi)$, it has 
been argued in Appendix~\ref{compo} that the active terms continue to dominate 
in the hydrodynamic limit and thus our results still hold. For a 
tensionless lipid membrane that undergoes only a second order MPT in 
equilibrium, $\sigma(\phi)=0$ identically.

\section{Nature of MPTs}\label{mpts}

{{There is no general framework available to study phase 
transitions in nonequilibrium systems. Nonetheless we are able to analyse the 
MPTs here by drawing analogy between the effective theory for composition 
fluctuations and standard equilibrium results. }}
 
{{Regardless of the nature of MPTs,
we have for a symmetric membrane $\langle\nabla^2 h\rangle=C=0$. Now, 
ignoring height fluctuations, if we substitute $\nabla^2 h$ by $C=0$ in 
(\ref{phieq}),
it reduces to the standard equilibrium, conserved dynamics (model 
B in the nomenclature of Ref.~\cite{halpin}), corresponding to the 
free energy (\ref{free-en}) [with $h=const.$]  that yields phase behaviour 
identical to the 
standard liquid gas transition at temperature $D_\phi$: phase 
coexistence for $r<0$ or $T<T_c$ with an asymmetric coexistence curve about 
the critical density and a critical point at $T=T_c$
 belonging to the $2d$ Ising universality class~\cite{chaikin}. Thus any 
active modification of this equilibrium-like picture should be fluctuation 
induced. To investigate that further, we follow Ref.~\cite{john2} 
and perturbatively integrate out 
$h$-fluctuations and obtain an effective dynamical equation for $\phi$ only. }}
Operationally, 
to account for the fluctuation effects at the 
simplest level, we calculate the effective (fluctuation-corrected) parameters of 
(\ref{phieq}) at the lowest order {in perturbative expansions}. We then 
express (\ref{phieq}) in terms of 
the fluctuation-corrected parameters and substitute $\nabla^2 h$ by
$\langle\nabla^2 h\rangle (=0)$, ignoring fluctuations. In this approximation, 
the $\phi$-dynamics is entirely described by a fluctuation-corrected free 
energy $F_\phi$, {entirely decoupled from $h$} which has the form
{{
\begin{equation}
 F_\phi=\int d^dx [\frac{r}{2}\phi^2 + \frac{g_e}{3}\phi^3 + 
\frac{u_e}{4}\phi^4 + \frac{v}{6}\phi^6-\tilde h \phi],\label{fphieq}
\end{equation}
with
\begin{equation}
 \frac{\partial\phi}{\partial t}=\Gamma_\phi\nabla^2\frac{\partial 
F_\phi}{\partial \phi} + {\boldsymbol\nabla}\cdot {\bf f}_\phi\label{effphieq}
\end{equation}
as the attendant effective dynamical equation for $\phi$.
In (\ref{fphieq}), $u_e$ and $g_e$ include one loop corrections to the bare couplings $u$ and $g$, respectively. Here
 we have ignored the corrections to $r$ and $v$ 
since these are not central to the analysis here; a 
correction to $r$ merely shifts $T_c$ and we assume $v$ is always positive. 
{Further, $\tilde h$ has no active corrections at the lowest 
order.} Thus the one-loop effective dynamics of $\phi$ follows the
equilibrium model B dynamics with free energy $F_\phi$ at an effective 
temperature $D_\phi$.}

%first order MPT  (SMPT) requires $u_e$, 
%the {\em effective} or {\em fluctuation
%corrected}
%$u$ in (\ref{phieq}) to be negative (positive) for first order MPT 
%(SMPT)~\cite{chaikin,otherfluc}.
 
%existing (bare)
%model parameters would necessarily originate from nonlinear fluctuation effects
%and can be
%analyzed in perturbative expansions in powers of the nonlinear coupling
%constants. %In general, when
%\begin{eqnarray}
 %u_e  &<& 0 \implies {\rm first~ order}\nonumber \\
 %    &=& 0 \implies {\rm tricritical~ point ~ (TCP)}\nonumber \\
 %    &>& 0 \implies {\rm second~ order}
%\end{eqnarray}
%In this model, there are fluctuation corrections to $u$ that vanish when $u=0$
%and then there are corrections which {\em do not} vanish even when $u=0$. For a
%given $u$ the latter contributions may be tuned to make $u_e$ positive or
%negative, giving SOMPT or FOMPT, respectively.

With $\lambda>0$, retaining only the inhomogeneous {\em 
active}
one-loop
correction to $u$ and $g$ (this suffices 
for our 
arguments here)
%\begin{equation}
 %$\Delta u= \frac{18 \lambda_2 \tilde \lambda^3}{3!}
 %\int^\Lambda \frac{d^2q}{(2\pi)2}\frac{d\Omega}{2\pi}
%(-q^6)C_{hh}(q,\Omega)G_h(q,\Omega)^3,$
%\end{equation}
%such that $u_e=u+\Delta u$, where $\Omega$ is a frequency and $G_h
%(q,\Omega) (=\langle \delta h/\delta
%f_h\rangle)$ is the propagator of $h({\bf
%q},\Omega)$ and $C_{hh}=\langle 
%|h({\bf q},\Omega)|^2\rangle$ is the height-height
%correlator. $\Lambda$ is an upper wavevector cut off; see SM for 
%details. 
%As we apporach the FOMPT temprature, say $T^*$
%from above, $\langle\phi\rangle=0$ but $\langle\phi^2\rangle= \phi_0^2$ is a
%finite constant (no large fluctuation of $\phi$
%at $T^*$). Hence, Eq.~(\ref{heq}) takes a simpler linear form
%\begin{equation}
% \frac{1}{\Gamma_h} \frac{\partial h}{\partial t} = -\kappa q^4 h - \lambda
%\phi_0^2 q^2 h + \eta_h,\label{heq1}
%\end{equation}
%yielding $G_h=1/[-i\omega + \Gamma_h(\kappa q^4 +\lambda \phi_0^2 q^2)]$ and
%$\langle |h({\bf q},\omega)|^2\rangle = 2D_h\Gamma_h/[\omega^2
%+\Gamma_h^2(\kappa q^4 +\lambda \phi_0^2 q^2)^2]$.
%In particular, %taking effective $C_{hh}(q,\Omega)=2D_h\Gamma_h/[\Omega^2 +
%\Gamma_h^2\kappa_e^2 (q) q^8]$ and effective  $G_h(q,\Omega)=1/[-i\Omega + 
%\Gamma_h
%\kappa_e(q)q^4]$ (which now include active tensions) 
%with the forms for $G_h$ and $C_{hh}$ in SM,
we obtain
%\begin{eqnarray}
$ u_e = u+\Delta u = u+ \lambda_p \tilde \lambda^2 D_h A$ and $g_e=g+ 
\tilde\lambda^2 \lambda_2 D_h B_1 + \lambda\lambda_2 D_h B_2$
%\end{eqnarray}
%{\bf Check if this expression is for with or w/o hydrodynamics}
where, $\lambda_p=\lambda_2 \tilde\lambda$ and $A, B_1, B_2$ are numerical 
constants; 
{see Appendix C}.  Thus effective couplings $u_e$ and $g_e$
can be independently positive, negative or zero. 

{
The phase behaviour and transitions of $\phi$ can be directly obtained from 
$F_\phi$. 
First consider $u_e>0$. The term $v\phi^6$ in (\ref{fphieq}) 
is now redundant and we ignore it. Then $F_\phi$ has the same form as 
$\mathcal F$ in (\ref{free-en}). As a result, the discussions that 
immediately follow $\mathcal F$ apply to $F_\phi$ as well: by making a 
suitable shift in $\phi$, the cubic term $g_e\phi^3$ may be eliminated from 
$F_\phi$, yielding a modified form for $F_\phi$ identical to the free energy 
for the Ising model in the 
presence of an external magnetic field $\tilde h_\phi$. Then, composition 
$\phi$ 
generally undergoes a first order MPT below a transition temperature.  A 
critical point 
with a second order MPT may be accessed only by suitable tuning of {\em 
both} $T$ and 
$\tilde h_\phi$: 
in fact the critical point is located in the $(T, \tilde h_\phi)$ plane at 
$r=0$ or 
$T=T_c$ and $\tilde h_\phi=0$, with an associated universal scaling behaviour 
belonging to the $2d$ Ising universality class~\cite{chaikin}. 
 Since $\tilde h_\phi$ in general does depend on the active coefficients $\lambda,\tilde \lambda$, 
 tuning activity can make $\tilde h_\phi =0$. Further, $T_c$ 
 too receives fluctuation corrections that depends on activity (not shown here). 
 Thus the critical point for this nonequilibrium second order MPT can be 
 accessed by controlling the activity. The role of 
$g_e\neq 0$ is only
to introduce an asymmetry of the order parameter $\langle\phi\rangle = m$ about 
the critical density, 
reflected in the curvature of the coexistence curve at the 
criticality~\cite{chaikin}. Since $g_e$ can be varied continuously and made 
positive, negative or zero by tuning the composition-membrane
interactions parameters, the curvature at 
criticality and hence the location of the coexistence curve in the 
$\phi-D_\phi$ plane changes continuously with the active parameters.
Experimental measurements of the coexistence curve for a given system can thus 
reveal valuable quantitative information about the active coefficients and the 
underlying active processes in the membrane.}

To ensure thermodynamic stability for $u_e<0$, we need to include the 
$v\phi^6$-term into consideration. This then yields a
first order MPT at
temperature $T^*= T_c+2u_e^2/(3v)$ and 
$m^2=|u_e|/(2v)$ even for $\tilde h_\phi=0$ in direct analogy 
with the known equilibrium MF results~\cite{chaikin}. 
 At
 the tricritical point, $u_e=0$, i.e., $u= - \lambda_p \tilde \lambda^2 D_h n 
A$ and $\tilde h_\phi=0$. Notice that,
unlike
equilibrium examples of tricritical points~\cite{chaikin}, here the condition 
for the tricritical point explicitly
involves  $D_h$, thus
bearing the hallmark of nonequilibrium~\cite{noneqTP}. 
{While the MPTs for $u_e>0$ are essentially indistinguishable 
from their equilibrium counterparts or the equilibrium liquid-gas phase 
transitions with a second order MPT accessible by setting $\tilde h_\phi=0$ and 
tuning $T$ 
to $T_c$, the prospect of a first order MPT for $u_e<0$ at $\tilde h_\phi=0$ at 
$T=T^*>T_c$ 
and the associated tricritical point 
are truly remarkable in that they have no analogues in the equilibrium limit of 
the MPTs or in the equilibrium
liquid-gas phase transition.
 Fluctuation induced shifts in $T_c$ and $T^*$ 
due to $\lambda_p\neq 0$ are argued to be
finite ({see Appendix}), suggesting $T_c$ and $T^*$ to 
be 
experimentally accessible at least
for certain inversion-symmetric mixed membranes with proper choices for the 
model 
parameters.  
 A
schematic phase diagram  of the
model in the
$\lambda_p-u$ plane with $\lambda>0$ and $\tilde h_\phi=0$) is shown in 
Fig.~\ref{phase}.
\begin{figure}[htb]
 \includegraphics[width=9cm]{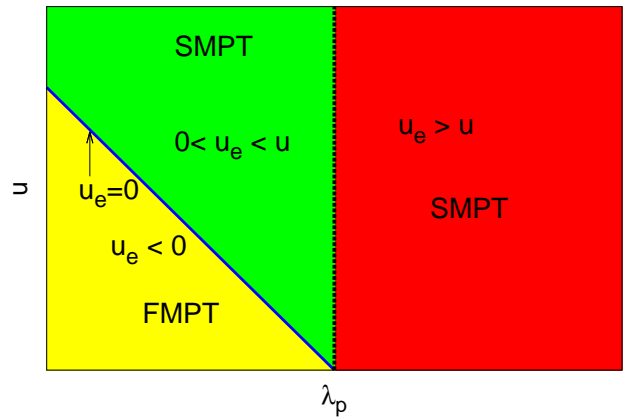}
\caption{(Color online) Schematic phase diagram in the 
$\lambda_p-u$ plane ($\lambda,u>0,\tilde h_\phi=0$). Inclined blue line marks 
the line of 
tricritical points (boundary between first order MPT and second order MPT : 
$u_e=0$; see text). 
The vertical broken line
(separating the red and green regions) represents $u_e=u$.} 
\label{phase}
\end{figure}
Our analysis of MPTs are only indicative in nature that may be confirmed by 
detailed numerical studies.
%numerical studies should be performed to conclusively settle the nature of 
%the MPTs; see Ref.~\cite{recent} for numerical studies on related systems.
%numerical studies should provide definitive confirmation of our 
%analysis of MPTs above.

\section{Summary and outlook}\label{summary}

We have thus developed an active hydrodynamic theory for
inversion-symmetric  mixed membranes. { There may be a variety of physical 
realisations which may be described by our model in the hydrodynamic limit, 
e.g., a mixed symmetric lipid membrane immersed in an isotropic active fluid or 
a a mixed symmetric lipid membrane with an active component in a passive fluid.}
We  
demonstrate that the interplay between
heterogeneity and the active processes  in the form of lipid-dependent 
active tensions leads to nontrivial fluctuation 
properties of mixed membranes. {We 
establish a direct correspondence 
between membrane conformation fluctuations and 
MPTs in mixed lipid membranes, which forms a key result of this work. This 
can be tested in {\em in-vitro} 
experiments on various physical realizations of our model; membrane 
fluctuations may be measured by spectroscopic 
methods~\cite{timo}. { In particular, 
tagged particle diffusion measurements~\cite{ref} may be used to validate our 
results.  We 
welcome construction of 
lattice-gas type nonequilibrium models which will be particularly suitable to 
numerically study and verify the results obtained here; see, e.g., 
Ref.~\cite{machta}. } 
 Our work clearly provides a way to ascertain the sign of $\lambda$ and the 
nature of MPTs {\em 
without} measuring $\phi$-fluctuations.} For a jump in $\sigma_a$ or $\kappa_e$ 
at a given 
$T=T^*$, it must be first order MPT; else if $\sigma_a$ or $\kappa_e(q)$ rises 
smoothly 
and 
diverges at some $T=T_c$ as $q\rightarrow 0$, the system displays second order 
MPT. 
Furthermore, if $\sigma_a|_{disorder} <(>) 
\sigma_a|_{order}$, or equivalently, $\kappa_e|_{disorder} 
<(>)\kappa_e|_{order}$, then $\lambda > 
(<)0$.  A mixed lipid membrane immersed in an active isotropic fluid made 
of 
actin filaments is a possible {\em in vitro} system to study our theory. This 
may be possible by reconstituted actomyosin arrays on a liposome; see, e.g.,  
Ref.~\cite{murrell} for a related experimental study. ATP depletion 
methods~\cite{atp-dep} can be used to control the magnitude of 
$\lambda$ and $\tilde\lambda$.  It would be interesting to see how contractile 
or extensile active 
fluids~\cite{sriram-RMP} affect the couplings $\lambda,\tilde\lambda$. 
 The sign of 
$\lambda_p$, crucial in our theory for fixing the 
order of MPT, {may be varied by using different sets of lipids}. {It may be 
noted that in the ordered phase separated state with A- and B-rich domains, one 
may define 
  a  $\kappa_e$ in a given domain that is A- or B-rich. 
Measurement of this domain-dependent $\kappa_e$ can further yield 
information about both 
$\lambda,\tilde\lambda$; {see Appendix}.

  So far in the above, we ignored hydrodynamic 
friction~\cite{kremers}. Even when that is included in our analysis, our general 
conclusion of a one-to-one 
correspondence between 
the membrane conformation fluctuations and the nature of the MPTs holds 
good. Interestingly with hydrodynamic friction, our results hold true even if 
the membrane is {\em impermeable}, i.e., $v_{perm}=0$; {see 
Appendix}.   
{We have neglected the geometric 
nonlinearities in our analysis above. These originate from the nonlinear 
forms of the area element and mean curvature in the Monge gauge; see 
Ref.~~\cite{wein}.
These are irrelevant 
near the critical point in a scaling sense in the presence of the existing 
nonlinearities.
Across first order MPT these may
affect the first order transition temperature $T^*$ and 
order 
parameter $m$ quantitatively; however, our general conclusions are
expected to remain unchanged.} We did not include any $\phi^3$ 
or $\phi^5$ term in (\ref{free-en}) above, as we considered only second order 
MPT belonging 
to the Ising universality class
in the equilibrium limit. A $\phi^3$-term in (\ref{free-en}) above would 
yield an first order MPT in equilibrium; a $\phi^5$-term
would be irrelevant in a scaling sense in the presence of the $\phi^4$-term in 
(\ref{free-en}) above. Beyond its immediate applicability to 
suitable {\em in-vitro} systems, our theory 
should serve as a basis for studying the physics of phase transitions in live 
cell membranes.
We expect 
our theory to introduce new directions
in the physical understanding of living cell membrane dynamics  with new vistas 
of striking nonequilibrium phenomena. We look forward to experimental tests of 
our predictions on GPMVs and live cell membrane extracts.

%\begin{narrowtext}
%\begin{table}
 %\begin{center}
  %\begin{tabular}{ |p{0.8cm}|p{1.9cm}|p{2.5cm}|p{2.7cm}| }
  %\hline
  %$\lambda$ & Stiff/soft& FMPT at $T=T^*$ & SMPT at $T=T_c$\\
  %\hline
  %$\lambda>0$ & Generic stiff at all $T$ &Jumps in $\kappa_e(q),
 % \Delta_n$ finite $\Rightarrow$ orientational LRO, $\Delta_h\sim \ln 
%L\Rightarrow$
 % positional QLRO & $\kappa_e(q)$ rises smoothly as $T\rightarrow T_c$, 
%$\Delta_n$ smooth, finite at all $T\Rightarrow$ orientational LRO,
%$\Delta_h$  $\sim\ln L$ (smooth, QLRO) for $T\neq T_c$ and $\sim\ln\ln L$ 
%(NLO)  as $T\approx T_c$\\
%\hline
 %  $\lambda<0$ & Generic soft/ crumpling at all $T$& Finite $\zeta$, $\zeta 
%(T>T^*)>\zeta (T<T^*)$  & Finite $\zeta$, 
%$\zeta(T_c)\ll \zeta(T\neq T_c$) \\
%\hline
 % \end{tabular}
%\caption{Variation of $\kappa_e(q),\Delta_n,\Delta_h$ at 
%different $T$ for positive and negative $\lambda$.}
%\label{table}
 %\end{center}

%\end{table}

%\end{narrowtext}

%\paragraph{Acknowledgement:-} Discussions with A. Chakrabarti and 
%S. Ramaswamy are gratefully acknowledged.

\section{Glossary: $\Delta_n$, $\Delta_h$ and $\zeta$ for first order MPT and 
second order MPT}\label{glossary}

{{Below we provide a list of symbols and the principal results 
that form the basis of this work here.}}

\begin{equation}
 \Delta_h = \int_{2\pi/L}^{\Lambda} \frac{d^2q \, d\Omega}{(2\pi)^3} 
\langle|h(q,\Omega)|^2 \rangle,
\end{equation}
where $\Omega$ is a frequency and
$\langle |h(q,\Omega)|^2\rangle=\frac{\langle|f_h(q,\Omega)|^2 
\rangle}{\Omega^2+\Gamma_h^2\kappa_e^2 q^8}
=\frac{2 D_h \Gamma_h}{\Omega^2+\Gamma_h^2\kappa_e^2 q^8}$. Similarly,
\begin{equation}
 \Delta_n = \int_{2\pi/L}^{\Lambda} \frac{d^2q \, d\Omega}{(2\pi)^3} q^2 
\langle|h(q,\Omega)|^2\rangle
\end{equation}

We first assume $\lambda>0$.
The membrane becomes generically stiff for all $T$.
\par
{\bf Case I : first order MPT at $T=T^*,\; (\lambda_p<0$}; see main text). At 
all $T$
\par
 (i) $\Delta_n :$ finite. $\Rightarrow$  orientational LRO.
 \par
 (ii) $\Delta_h \sim {\rm ln}~L\Rightarrow$: positional QLRO.
\par
 (iii) Effective bending modulus, $\kappa_e$ shows a jump across $T^*$: 
$\kappa_e (T>T^*) < \kappa_e (T<T^*)$. 
\par
{\bf Case II: second order MPT at $T=T_c$.}
\par
For any $T \neq T_c$, 
\par
(i) $\Delta_n$ is finite and smooth (no jump across $T_c$)$\Rightarrow$: 
orientational LRO.
\par
(ii) $\Delta_h \sim {\rm ln}~L$ $\Rightarrow$: positional QLRO.
\par
(iii) At $T=T_c$, $\Delta_n:$ finite (orientational LRO), in fact further
suppressed than its value at $T \neq T_c$. On the other hand, $\Delta_h \sim 
{\ln \ln}~L$ at $T=T_c$,
$\Rightarrow$ positional NLO.
\par
(iv) Effective bending modulus,
$\kappa_e$ rises smoothly as $T_c$ is approached.
\par

Now assume $\lambda<0$: Generic crumpling is introduced at all $T$ giving a 
finite 
persistence length.
\par
(i)second order MPT: $\zeta (T<T_c)<\zeta (T>T_c)$; $\zeta (T=T_c)\ll 
\zeta(T\neq T_c)$.
$\zeta(T=T_c)$ follows the equation 
\begin{equation}
 \zeta^2 {\rm ln}(\zeta/2\pi) = \frac{8\pi^3\kappa}{D_{\phi}|\lambda|}.
\end{equation}

(ii) first order MPT: $\zeta (T<T^*) < \zeta (T>T^*)$.

\section{Acknowledgement}
The authors thank %A. Chakrabarti and S. Ramaswamy for discussions, and 
the Alexander von Humboldt Stiftung, Germany for partial financial support 
through the Research Group Linkage Programme (2016).

%\newpage

%\begin{center}
% \Large{\bf Appendix}
%\end{center}
\begin{appendix}
 
 \section{The active terms}
 
 We now briefly discuss a more formal derivation of the active 
terms. To this end, we closely follow Ref.~\cite{iso}, keeping the example in 
mind of a lipid membrane embedded in an isotropic active fluid 
made of actin filaments and motors. In the Monge gauge for 
a nearly 
planar 
membrane along the $xy$-plane, we can write for the local membrane velocity 
$v_z$ as 
appropriate for a symmetric membrane (set $\sigma=0$)
\begin{equation}
 \frac{\partial h}{\partial t}=v_z=v_{hydro_z}  +X(\phi)\nabla_j (p_z 
p_j)-\Gamma_h\frac{\delta {\mathcal F}}{\delta h}.
 \end{equation}
We use $p_i=\partial_i h,\,i=x,y$ at the location of the membrane; 
$v_{hydro_z}$ is the $z$-component of the three-dimensional hydrodynamic 
velocity ${\bf v}_{hydro}$. 
 This
 may be formally 
justified by closely 
following the arguments outlined in Ref.~\cite{iso}. We assume 
that the free 
energy is dissipated at a rate ${\mathcal R}\Delta\mu$, where ${\mathcal R}$ is 
the reaction rate of ATP hydrolysis and $\Delta\mu$ is as given in the main 
text. Now treat $v_z$ and $\mathcal R$ as 
fluxes, and $\delta {\mathcal F}/\delta h$ and $\Delta\mu$ as the corresponding 
conjugate thermodynamic forces. Then as in Ref.~\cite{iso}, together 
with the 
condition for a symmetric membrane, we identify $\overline\zeta\nabla_j(p_z 
p_j)\delta{\mathcal F}/\delta h$ as the leading order contribution to $\mathcal 
R$ up to the 
first order in gradients, where $\overline\zeta$ is an Onsager
coefficient. Using the symmetry of the dissipative Onsager 
coefficients, we then set $\overline\zeta \nabla_j(p_z 
p_j)=X(\phi)\nabla_j (p_z p_j) $, giving a justification of the active terms. 
%[{\bf SHOULD THIS BE REALLY HERE OR IN THE APPENDIX?}]

\section{Effective bending modulus}
In general, from Eq.~(\ref{heq}) ( after neglecting fluctuations 
$\delta X (\phi)$ w.r.t. the mean value  $\langle 
X(\phi)\rangle$), we obtain
\begin{equation}
\sigma_a = \lambda\langle\phi^2\rangle + \tilde \lambda 
\langle\phi\rangle,\;{\rm or,}\;
 \kappa_e (q) = \kappa + \frac{\lambda}{q^2} \langle \phi^2 \rangle + 
\frac{\tilde\lambda}{q^2}\langle\phi\rangle.
%\label{kappaeff}\\
\end{equation}
We have
$\langle\phi\rangle=0$, considering the whole system with equal amount of 
A and B lipids in the system. Now, the ordered phase is 
characterised by finite (macroscopic) size domains, which are A-rich or B-rich, 
for
$T<T_c$ (second order MPT) or $T<T^*$ (first order MPT). This allows us to 
define
  $\sigma_a$ or $\kappa_e$ over a single (macroscopic size) domain, either A or 
B 
rich with 
average of $\phi$ in a given type of domain being non-zero; $\sigma_a$ (or, 
$\kappa_e$) will now
depend explicitly on the domain type.  
We now define domain-dependent active tensions $\sigma_a^A$ and 
$\sigma_a^B$ for A- and B-rich domains in the ordered phase and find
\begin{equation}
 \sigma_a^A=\lambda\langle\phi^2\rangle_A + \tilde\lambda 
\langle\phi\rangle_A,\;
 \sigma_a^B=\lambda\langle\phi^2\rangle_B + \tilde \lambda  
\langle\phi\rangle_B.
\end{equation}
Here, $\langle...\rangle_{A,B}$ represents averages taken in an A or B rich 
domain, respectively. For simplicity, let us 
consider just two 
macroscopic size domains, one A-rich and another B-rich. Within a simple 
mean-field 
like description, we define an A (B) rich domain formally by 
$\langle\phi\rangle_A=m >0
$ ($\langle\phi\rangle_B = -m<0$) and set  $\tilde\lambda 
>0,\,\lambda>0$.
Evidently, for sufficiently large $\tilde\lambda$, $\sigma_a^B<0$. One may 
then
define a threshold $\tilde\lambda_c$, given by $\sigma_a^B=0$.  This yields 
that a 
 crumpling instability takes place in the B-rich domain for $\tilde\lambda > 
\tilde\lambda_c$ for a given size of the B-rich domain, where as $\sigma_a$ 
remains 
positive in the A-rich domain  and thus, the latter should be statistically 
flat. This 
is testable 
in experiments. In contrast, with $\lambda<0$, the crumpling instability 
in the A-rich domain in the ordered phase may be suppressed by a sufficiently 
large $\tilde \lambda$, such that $\sigma_a^A>0$ even with $\lambda<0$. This 
corresponds to a novel situation, where a large enough flat mixed membrane as a 
whole is unstable in the 
disordered phase (since $\lambda <0$), but a macroscopic part of it (i.e., the 
A-rich domain) gets stabilised and shows statistical flatness in the ordered 
phase (large $\tilde \lambda>0$). Again, this should be testable in standard 
experiments. Overall we conclude that the stability and flatness of the mixed
membrane, both in the disordered and ordered phases, depend very sensitively on 
the active processes. 
Since there are no domains in the disordered phase, the term with coefficient 
$\tilde\lambda$ has no effect on $\sigma_a$ in the 
disordered phase, independent of first order MPT or second order MPT. Thus, 
the sign of $\sigma_a$ is necessarily controlled by $\lambda$ in the 
disordered phase  and our results in main text directly apply. 
For sufficiently small $\tilde \lambda$, note that the results do not change
qualitatively, and Eq.~(\ref{activeten}) remains valid in each domain.

\section{Analysis of the MPTs : Active inhomogeneous fluctuation corrections to 
$u$ and $g$}

We begin with the generating functional~\cite{janssen} 
${\mathcal Z}$ corresponding to the Eqs. of motion (\ref{heq}) and 
(\ref{phieq}). We find
\begin{equation}
 {\mathcal Z}= \int {\mathcal D}h{\mathcal D}\hat h{\mathcal D}\phi {\mathcal 
D}\hat\phi \exp (S),
\end{equation}
where the action functional 
\begin{widetext}
\begin{eqnarray}
S &=& \int d^dx dt [\frac{D_h}{\Gamma_h} \hat h \hat h- \hat h(\frac{1}{\Gamma_h}\frac{\partial h}{\partial t}+\kappa \nabla^4 h -
\lambda \phi^2 \nabla^2 h - \tilde\lambda \phi \nabla^2 h) + \hat \phi (\frac{-D_\phi \nabla^2}{\Gamma_\phi}) \hat \phi
 -\hat \phi (\frac{1}{\Gamma_\phi}\frac{\partial \phi}{\partial t} \\ \nonumber
 &-& \nabla^2[r\phi-\nabla^2 \phi+g\phi^2+\frac{u}{3!}\phi^3+2\lambda_1\phi(\nabla^2 h)^2+\lambda_2(\nabla^2 h)^2+v\phi^5])]
\end{eqnarray}
\end{widetext}
One can then formally integrate out $h$ and $\hat h$, and define an effective 
action functional ${S}_\phi$ as follows:
\begin{equation}
 \exp (S_\phi) = \int {\mathcal D}h {\mathcal D}\hat h \exp (S).
\end{equation}
To evaluate $S_\phi$, we proceed perturbatively and then extract 
$F_\phi$~\cite{john2}, such that 
the effective $\phi$-dynamics is now given by Eq.~(\ref{fphieq}). Ignoring the 
fluctuation corrections of $\Gamma_\phi$ and $D_\phi$, as these are not central 
to the discussion here,

we find
\begin{equation}
 F_\phi = \int d^dx [\frac{r_e}{2}\phi^2 + 
\frac{1}{2}({\boldsymbol\nabla}\phi)^2 + \frac{g_e}\phi^3+\frac{u_e}{4}\phi^4 + 
\frac{v_e}{6}\phi^6],
\end{equation}
where a subscript $e$ refers to fluctuation-corrected parameters. 
The corresponding effective equation of motion of $\phi$ is
\begin{equation}
 \frac{\partial\phi}{\partial t}=\Gamma_\phi\nabla^2\frac{\delta 
F_\phi}{\delta\phi} + {\boldsymbol\nabla}\cdot {\bf f}_\phi.
\end{equation}
 In our analysis in the main text, we ignore the difference between $r_e$ 
and $r$, and between $v_e$ and $v$, as these are of no significance to the 
mean-field like arguments used in the main text. We always assume $v>0$. 
We also ignore the corrections to $D_\phi$, since that just changes the 
effective temperature and is of no direct consequences here. Furthermore, there 
are no one-loop corrections to $\Gamma_\phi$, owing to the conservation law form 
of (\ref{phieq}).

%\section{Active inhomogeneous fluctuation correction to $u$} \label{ue}
{To proceed further, we now need to find $g_e$ and $u_e$ perturbatively.}
The lowest order active inhomogeneous fluctuation corrections to $g$ and $u$ 
may be
represented by the following Feynman diagrams; see Figs.~\ref{g_e1},~\ref{g_e2} 
and~\ref{u_e} below.

\begin{figure}[htb]
 \includegraphics[width=8.6cm]{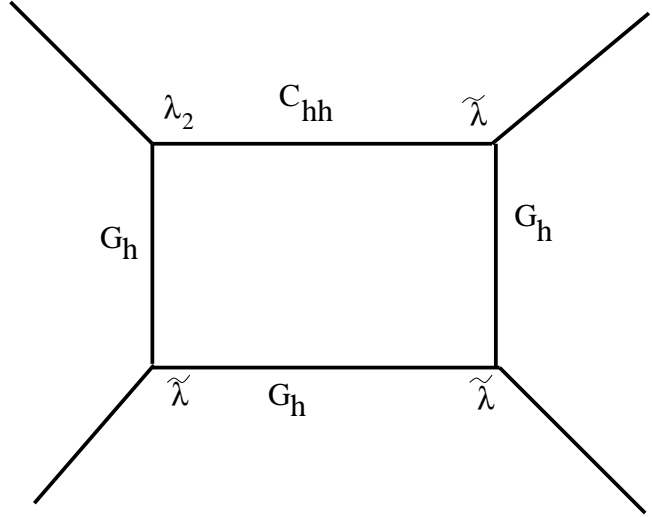}
\caption{Active fluctuation correction to $u$ due to 
$\tilde\lambda,\lambda_2$.}  
\label{u_e}
\end{figure}

\begin{figure}[htb]
\includegraphics[width=8.6cm]{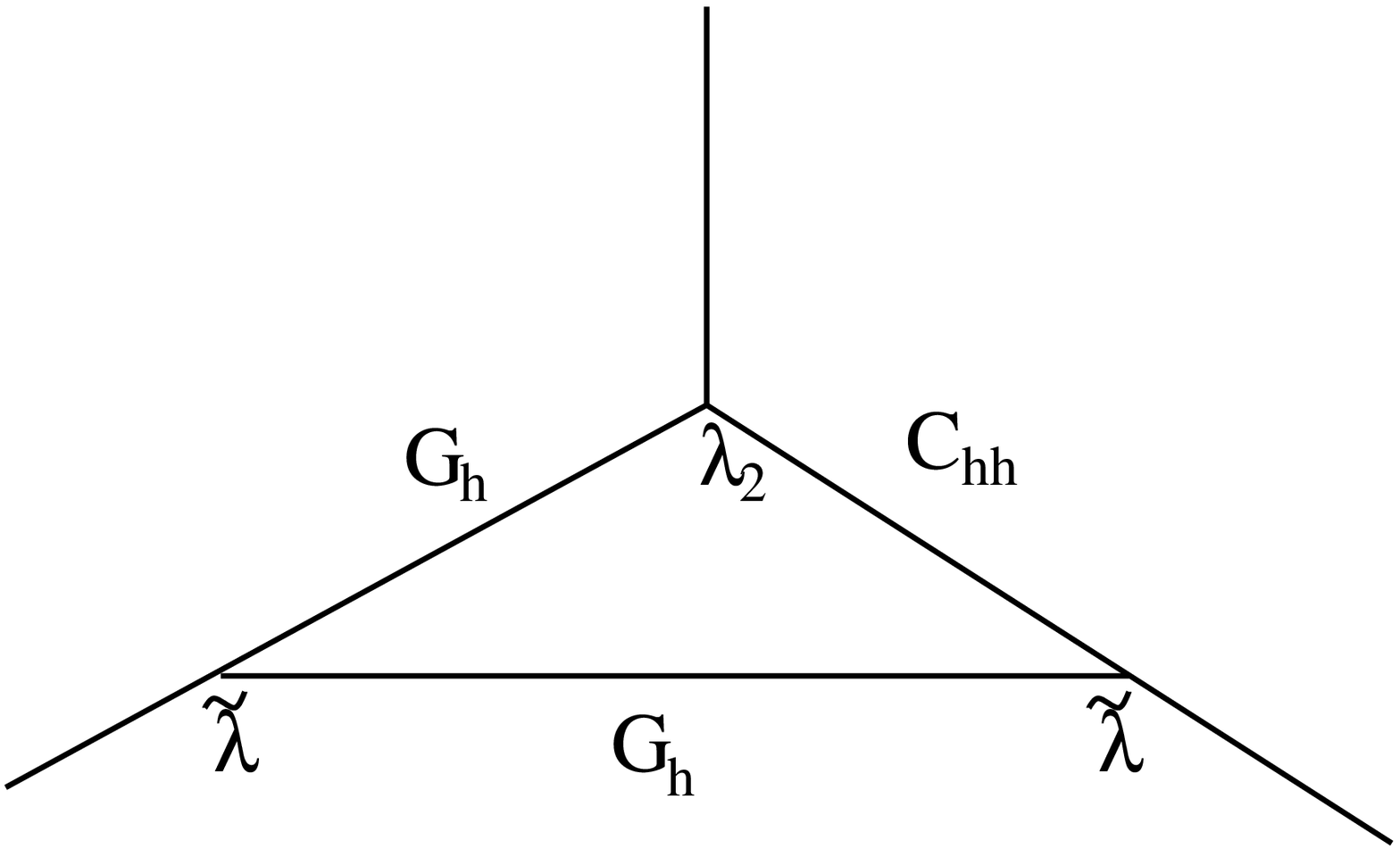}
\caption{Active fluctuation correction to $g$ due to 
$\tilde\lambda,\lambda_2$.}  
\label{g_e1}
\end{figure}

\begin{figure}[htb]
\includegraphics[width=8.6cm]{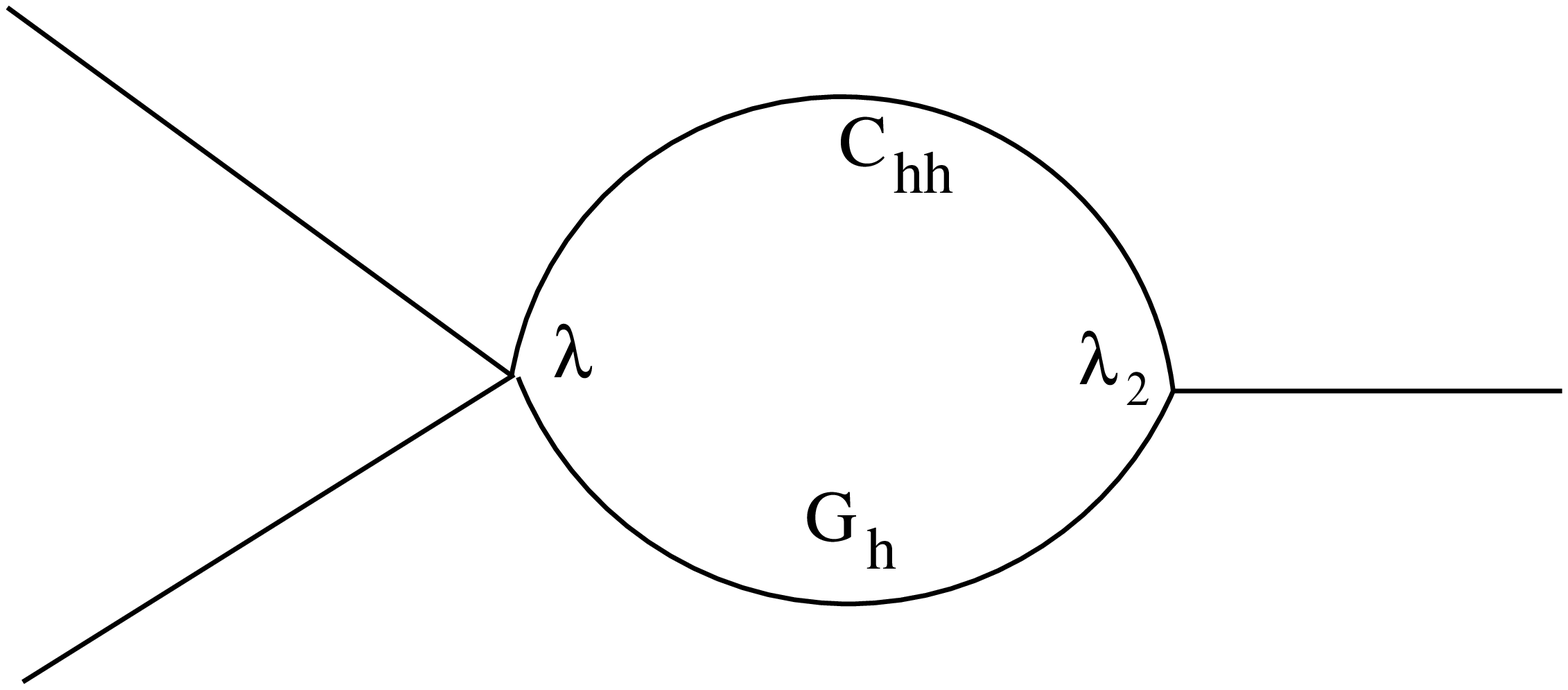}
\caption{Active fluctuation correction to $g$ due to 
$\lambda,\lambda_2$.}  
\label{g_e2}
\end{figure}

The correlator function for $h$, written in terms of the effective bending 
modulus $\kappa_e(q)$, is given by
$C_{hh} (q, \Omega)=\langle |h({\bf q},\Omega)^2| \rangle=\frac{2 
D_h \Gamma_h}{\Omega^2+\Gamma_h^2\kappa_e^2 q^8}$. In addition,
$G_\phi (q, \Omega)=\frac{1}{-i \Omega + \Gamma_\phi q^2 (r+q^2)}$ and
$G_h (q,\Omega)=\frac{1}{-i \Omega + \Gamma_h \kappa_e q^4}$ 
are the
propagators for $\phi$ and $h$, respectively. 

 Furthermore, we assume $\lambda>0$ for stability.  The one-loop 
correction, $\Delta u$~(see main text),
thus evaluates to the following at $2d$:
\begin{equation}
 \Delta u = \frac{3 \lambda_p \tilde \lambda^2 D_h}{4} 
\int_{2\pi/L}^{\Lambda}\frac{ d^2 q}{(2\pi)^2 \, \kappa_e^4 \, q^6} 
\end{equation}
Since $\kappa_e(q)\sim O(1/q^2)$ for sufficiently small $q$, $\Delta u$ is 
clearly
a finite contribution at TL, as argued for in the main text. 
Similarly, the one-loop correction, $\Delta g$ of $g$, 
at $2d$ evaluates to

\begin{equation}
 \Delta g= \frac{\tilde\lambda^2 \lambda_2 D_h}{2}B_1 + 2\lambda \lambda_2 
D_h B_2,
\end{equation}
where, $B_1=\int_{2\pi/L}^{\Lambda}\frac{d^2 q}{(2\pi)^2} \frac{1}{2\kappa_e^3 
q^4}$ and 
$\Delta_2=\int_{2\pi/L}^{\Lambda}\frac{d^2 q}{(2\pi)^2} \frac{2}{\kappa_e^2 
q^2}$ are finite in $2d$.
The sign of $\Delta u$  can thus be varied by varying $\lambda_p$; 
for sufficiently 
negative 
large $\lambda_p$, $u_e$ can be made negative. Similarly, the sign of $\Delta 
g$ may be varied and may be made positive, negative or zero by tuning 
$\lambda_2$.

\section{DRG flow equations and fixed points}
Large critical point fluctuations very close to second order MPT  may
be
systematically handled within dynamic renormalization group (DRG)
frameworks; see Ref.~\cite{chaikin} for technical details. With 
$\tilde\lambda=0=\lambda_2$, simple 
power counting 
shows that the
nonlinear
coefficients $\lambda$ and $u$ are {\em equally relevant} (in a scaling/DRG
sense) at $2d$, the physically relevant dimension, with both being
{\em marginal} at $d=4$. This calls for a perturbative DRG calculation to be
performed on Eqs.~(\ref{heq}) and (\ref{phieq}), together
with an $\epsilon$-expansion, $\epsilon=4-d$; see Ref.~\cite{chaikin}.
In 
this limit, the  model admits only second order MPT belonging to the $2d$ Ising 
universality 
class. The one-loop DRG procedure formally involves the following steps (i) 
obtaining the one-loop fluctuation corrections to the different model 
parameters by 
integrating out the high wavevector parts of the fields from $\Lambda/b$ to 
$\Lambda$, $b>1$, (ii) rescaling the fields wavevector $\bf q$, frequency 
$\omega$ by ${\bf q}'=b{\bf q},\,\omega'=b^z\omega$, where $z$ is the dynamic 
exponent and rescaling of the fields 
$h$ and $\phi$ accordingly~\cite{chaikin}. The fixed 
points (FP) of the DRG are to be obtained from the flow equations for the
relevant coupling constants in the problem, which in the present case are 
$u$ and $\lambda$. The one-loop diagrams that contribute to $u$ and $\lambda$ 
are shown below
in Fig.~\ref{u_new_dia} and Fig.~\ref{lambda_new_dia}, respectively.

\begin{figure}[htb]
\includegraphics[width=8.6cm]{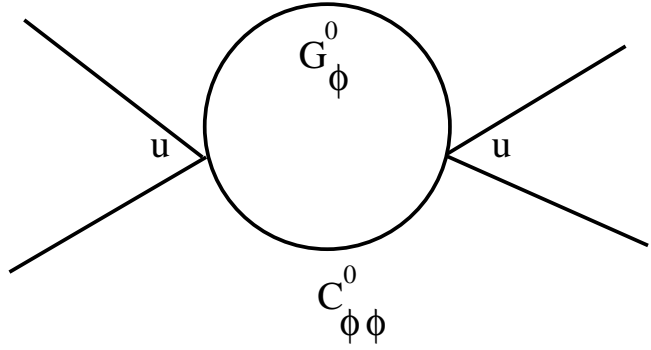}
\caption{Relevant one-loop correction to $u$. Here, $G_\phi^0=\frac{1}{-i 
\Omega + \Gamma_\phi q^2 (r+q^2)}$
and $C_{\phi\phi}^0=\frac{2D_\phi \Gamma_\phi}{\Omega^2+\Gamma_\phi^2 q^4 
(r+q^2)^2}$ are 
the bare propagator and correlator of $\phi$, respectively.} \label{u_new_dia}
\end{figure}

\begin{figure}[htb]
\includegraphics[height=8cm]{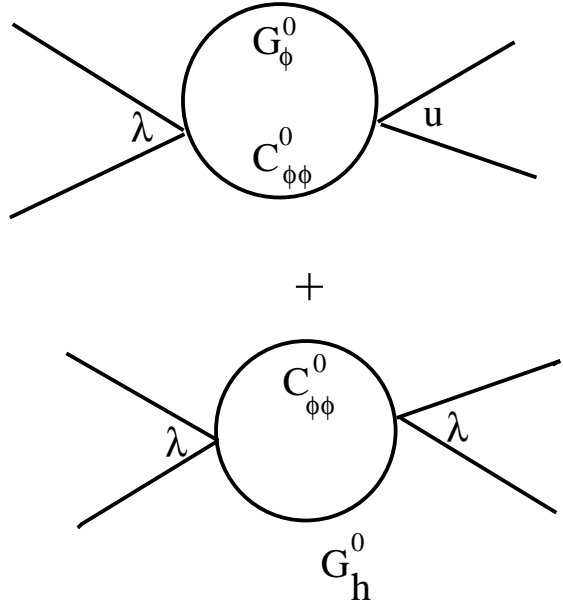}
\caption{Total correction to $\lambda$. Here, $G_h^0=\frac{1}{-i \Omega + 
\Gamma_h \kappa q^4}$ is
the bare propagator for $h$; $C^0_{\phi\phi}$ and 
$G^0_\phi$ are as in Fig.~\ref{u_new_dia}.}\label{lambda_new_dia}
\end{figure}

With $b=e^l 
\approx 1+l,\,l\ll 1$,
the resulting DRG recursion relations for
 $u$ and $\lambda$ yield %are
\begin{eqnarray}
 \frac{du}{dl}&=&u[\epsilon-9uD_\phi],\label{uflow}\\ 
\frac{d\lambda}{dl}&=&\lambda[\epsilon-3uD_\phi-4\tilde \Gamma \lambda 
D_\phi],\label{lambdaflow}
\end{eqnarray}
where $\tilde \Gamma=\frac{\Gamma_h}{\Gamma_\phi+\kappa \Gamma_h}$.
For the flow equations~(\ref{uflow}) and (\ref{lambdaflow}), stable FP 
 $u^*=\frac{\epsilon}{9D_\phi}$ and 
$\lambda^*=\frac{\epsilon}{6\tilde \Gamma D_\phi}$.
 Not surprisingly, $u^*=\epsilon/(9D_\phi)$ yields the critical exponents for 
the composition fluctuations at second order MPT identical to
their values at the Heisenberg FP of the Ising model in equilibrium, consistent
with the expectation that the second order MPT belongs to the Ising universality 
class. %
At the
one-loop order, there are no fluctuation corrections to
$\Gamma_h,\Gamma_\phi,D_h$ and $D_\phi$. Now, since we are 
formally considering a tensionless membrane, noting that the leading order 
correction to the self-energy of $h({\bf q},\omega)$ is at $O(q^2)$, it is 
convenient to define $\tilde\kappa(q)=\kappa q^2$ ($\tilde\kappa$ clearly has 
the physical dimension of a surface tension), and obtain its 
fluctuation-corrections. We find
\begin{equation}
 \frac{d\tilde\kappa}{dl}=\frac{\lambda D_\phi}{2\pi}
\end{equation}
as the DRG flow equation for $\tilde\kappa$. Solving this at the DRG FP, we 
obtain a scale-dependent, renormalised $\tilde\kappa(q)$ and thence, 
defining renormalised $\kappa_e(q)$ via $\tilde\kappa(q) = \kappa_e(q) 
q^2$,
\begin{equation}
 \kappa_e(q)= \kappa+\frac{\lambda^* 
D_\phi}{q^2}\int_q^{\Lambda}\frac{d^2q_1}{(2\pi)^2q_1^2} \approx 
-\frac{\lambda^* D_\phi}{2 \pi q^2}{\rm ln} 
q ,\label{effkappa11}
\end{equation}
at $d=2$ for small $q$ at the DRG FP or the critical point ($r=0$). The 
one-loop correction
to $\kappa$ is shown in Fig.~\ref{kappa_new_dia}.

\begin{figure}[htb]
\includegraphics[height=6.8cm]{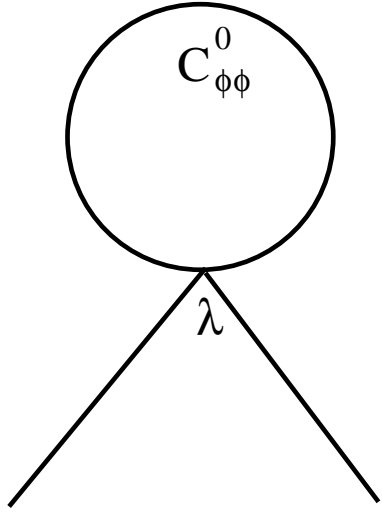}
\caption{One-loop correction to $\kappa$ that yields $\kappa_e(q)$.} 
\label{kappa_new_dia}
\end{figure}
%The values of $G_h, G_\phi$ and $C_\phi$ are as mentioned in Sec. XI of SM.

Here, at the DRG fixed point $\lambda^*D_\phi=\epsilon/(6\tilde\Gamma)$. 
 Then,
%\begin{equation}
$\Delta_n=-
 \frac{D_h}{\lambda^* D_\phi} \int^{\Lambda}_{q_0} \frac{q_1 dq_1}{{\ln}~q_1}$, 
where 
$q_0\sim 2\pi/L$ is the lower limit of the integral. The nature of 
the orientational 
order is determined by the $L$-dependence of $\Delta_n$ in the limit of large 
$L$ (formally infinite at TL). Whether or not 
the limit 
$q_0\rightarrow 0$ may be taken, depends on the behavior of the integrand, $q_1$ 
for 
$q_1\rightarrow 0$. It can be shown in a straight forward way that for 
$q_1\rightarrow 0$,  $q_1/\ln q_1$ vanishes. Thus, we conclude that for 
$q_0\sim 2\pi/L \rightarrow 0$, $\Delta_n$ remains finite. A precise numerical 
value of $\Delta_n$ may be obtained by numerical integrations. This, of course, 
will depend upon $\Lambda$, the upper limit. Since this is not particularly 
illuminating for the purposes of this work, we do not do this here.
Furthermore,
$\Delta_h=-\frac{2\pi D_h}{\lambda^* D_\phi}\int_{2\pi/L}^{\Lambda}\frac{d^2 
q_1}{(2\pi)^2 q_1^2 
{\rm ln}~q_1} 
\approx \frac{D_h}{\lambda^* D_\phi}\ln\ln L$ in TL. Clearly, the 
amplitude $D_h/(\lambda^* D_\phi)$ 
is nonuniversal at the DRG FP and these results are in agreement with 
results obtained in the main text. Lastly, the lack of renormalization of 
$\Gamma_h$ and 
$\Gamma_\phi$ at the
one-loop order implies that dynamic exponent $z=4$ at second order MPT for both 
$h$ and
$\phi$, respectively (strong dynamic scaling). 

Note that similar to 
$u,\lambda$ and $\kappa$, $r$ also receives fluctuation corrections, reflecting 
fluctuation-induced shift in $T_c$. Solving the DRG flow equation for $r$ 
yields the correlation length exponent; see, e.g., Ref.~\cite{epje4}. 
Since this is not central to main issue of this work, we do not discuss this 
here.

{ 
{In the above, we have worked up to the one-loop 
approximation. Notice however that the critical behaviour of $\phi$ follows the 
Ising universality class as elucidated above, and hence the equal-time 
correlator of $\phi$ is 
known {\em exactly} near the critical point from the exact solution of the 
$2d$ 
Ising model. We use this below to obtain the temperature-gradient of the 
renormalised bending modulus $\kappa_e(q)$ near the critical point $T=T_c$ 
(see also Refs~\cite{john2,aronovitz}). 
From Eq.~(\ref{effkappa1}), only the equal-time 
correlator $\langle\phi^2\rangle$ enters into 
$\kappa_e$, giving~\cite{aronovitz}
\begin{equation}
 \frac{\partial \kappa_e}{\partial T}=\lambda\frac{\partial}{\partial T} 
\langle\phi^2\rangle\sim -C_v
\end{equation}
near the critical point. For the $2d$ Ising model, $C_v\sim \ln (|T - 
T_c|/T_c)$ near $T_c$, yielding a logarithmic divergence~\cite{2dising}. This 
shows how 
$\kappa_e(q)$ diverges as $T\rightarrow T_c$.
 This in turn yields, upon integrating over temperature, $\kappa_e(q)$ has a 
diverging piece $\propto \ln (|T-T_c|/T_c)$ near the critical point. Now 
noting that that correlation length $\zeta_c\propto (|T-T_c|/T_c)^{-\nu}$ near 
$T_c$ and setting $\zeta_c\sim 2\pi/q$ for the long wavelength modes, we find 
$\kappa_e(q)\propto \lambda q^2\ln q$ in the long wavelength limit, in 
agreement with our one-loop result above.}}

In the above, although we have neglected
 $\lambda_2$ and $\tilde\lambda$, the effective
coupling $\lambda_p=\tilde\lambda\lambda_2$ becomes marginal at $D=4$,
and hence should be {\em equally relevant} as $u$ and $\lambda$ in a DRG
sense. Indeed, there are additional one-loop corrections to the various bare
model parameters that originate from $\lambda_p$ (not shown here). 
Nonetheless, as our one-loop effective free energy $F_\phi$ suggests, the 
critical behaviour of $\phi$-fluctuations should still belong to the $2d$ Ising 
universality class. Thus, proceeding as above the divergence of $\kappa_e(q)$ 
near $T_c$ remains unchanged.

%We ignore this issue here.

\section{Fluctuation induced shift in $T_c$}

We now heuristically argue in favour of experimental accessibility of second 
order MPT and 
first order MPT in the system. Apart from the well-known shift in $T_c$ due to 
the $u$-term 
in 
Eq.(\ref{phieq})~\cite{chaikin}, there is a 
correction to $T_c$ of the form 
$\lambda_1\langle(\nabla^2 h)^2\rangle$, which is obviously finite. 
Additionally, 
a nonzero $\lambda_p$ should lead to a
fluctuation-induced shift in $T_c$. The corresponding one-loop Feynman diagram 
is shown 
in Fig.~\ref{newT_c} below.
 \begin{figure}[htb]
\includegraphics[width=8.6cm]{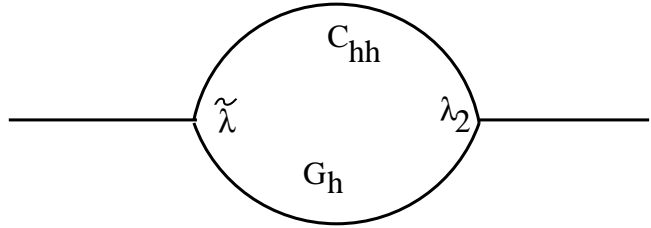}
\caption{Correction to $T_c$ coming from $\lambda_p$. Clearly, this contribution
being finite only produces a small shift.}  \label{newT_c}
\end{figure}
The expression is of the form $\sim \lambda_p D_h \int^{\Lambda}_{2\pi/L} 
\frac{d^2 q}{(2\pi)^2 
\kappa_e^2 q^2} $, which is finite. Thus, the shift in the mean-field $T_c$ due 
to 
the active effects is finite.
 This leads us to speculate that 
renormalised $T_{cR}$, i.e., the 
shifted or 
fluctuation-corrected
$T_c$, should fall in a temperature range similar to the equilibrium critical 
points 
of model lipid bilayers, and correspondingly, any putative second order MPT 
should be 
accessible in 
experiments, at least for certain choices of the parameters of the model 
system.
Furthermore, if we construct an effective Landau MF in terms of $u_e$ and
$T_{cR}$, the first order transition temperature $T^*$ also gets a shift. 
 Since 
$T_{cR}$ is expected to be experimentally accessible, the shifted $T^*$ should 
also be accessible experimentally for certain choices of the parameters of the 
model system. Thus, we speculate that it should be possible to 
observe MPTs (first order MPT or second order MPT) in certain 
inversion-symmetric 
lipid membranes, with properly tuned values of the model parameters, within 
experimentally 
accessible temperature ranges.

%{\bf NO $\gamma_\phi$ 
%defined in this section}
%Notice that if we consider $A$ to originate from a free energy, $A>0$ due to 
%reasons of thermodynamic stability. On the other hand, if $A$ is assumed to 
%originate from an active stress as above, then no restriction applies on its 
%sign.
%For a membrane with vanishing surface tension, $A=0=B$ 
%identically, and $\lambda,\tilde\lambda$ are the only active parameters in the 
%model.
%{\bf AB : When discussing EQUILIBRIUM surface tension, let us not mention 
%about 
%the active origins of $A$ and $B$. We are
%anyway talking about it in the next section. It reads better this way.}

\section{Composition-dependent surface tension}\label{compo}

We now briefly consider the effects of a composition-dependent surface 
tension $\sigma(\phi)$. For simplicity we choose $\sigma(\phi)= \lambda_3\phi^2 
+ \lambda_4\phi$ 
in the free energy $\mathcal{F}$ as given in (\ref{free-en}). For reasons of 
thermodynamic stability, we choose $\lambda_3>0$; the sign of 
$\lambda_4$ is arbitrary and may be absorbed in the definition of $\phi$. We 
choose the magnitude of $\lambda_4$ in a way to ensure $\sigma(\phi)>0$ for all 
$\phi$, again to ensure thermodynamic stability. This now yields
\begin{eqnarray}
 \frac{\partial h}{\partial t}&=& \Gamma_h [ (\lambda\phi^2 
+\tilde\lambda\phi)\nabla^2 h -\kappa\nabla^4 h  + 
\lambda_3{\boldsymbol\nabla}\cdot(\phi^2 {\boldsymbol\nabla}h) \nonumber \\ &+& 
\lambda_4
{\boldsymbol\nabla}\cdot(\phi {\boldsymbol\nabla}h)],\nonumber \\
\label{neweq3}
\frac{\partial\phi}{\partial t}&=&\Gamma_\phi 
\nabla^2[r\phi -\nabla^2\phi +
\frac{u}{3!} \phi^3 +2\lambda_1 \phi(\nabla^2 h)^2+g\phi^2 \nonumber \\&+& 
\lambda_2
(\nabla^2 h)^2+v\phi^5+ (2\lambda_3 \phi+ \lambda_4) \nabla^2 h]\nonumber \\ &+& {\boldsymbol\nabla}\cdot{\bf f}_\phi.
\end{eqnarray}
Thus, compared to Eq.~(\ref{heq}), there are additional terms with coefficients 
$\lambda_3,\lambda_4$ in Eq.~(\ref{neweq3}). Notice also that the term 
$\lambda_3{\boldsymbol\nabla}\cdot(\phi^2{\boldsymbol \nabla}) h$ has the same 
number of $\phi$ and $h$ fields as in the active $\lambda$-term in 
Eqs.~(\ref{heq}) or (\ref{neweq3}); similarly, the term 
$\lambda_3{\boldsymbol\nabla}\cdot(\phi{\boldsymbol \nabla}) h$ has the same 
number of $\phi$ and $h$ fields as in the active $\tilde\lambda$-term in 
Eqs.~(\ref{heq}) or (\ref{neweq3}).
 Still, these $\lambda_3-$ and $\lambda_4-$terms
are total derivatives, where as the active $\lambda$ 
and $\tilde\lambda$-terms are not. Thus in the hydrodynamic limit, the 
$\lambda_3$- 
and $\lambda_4$-terms may be neglected (in a scaling sense) in comparison with 
the $\lambda$ 
and $\tilde\lambda$-terms in (\ref{neweq3}). 
 It 
is important to note that if 
$\sigma(\phi)$-term is included in $\mathcal{F}$, then the upper 
critical dimensions~\cite{chaikin} of the $\lambda_3$- and $\lambda_4$-terms 
are 4 and 6, respectively. Thus, the MPT  of the $\phi$-fluctuations should no 
longer belong to the $2d$ Ising 
universality 
in the equilibrium limit. Given that in our work, we have 
considered 
a tensionless membrane that undergoes only second order MPT with 2d Ising 
universality at 
equilibrium, we set $\sigma (\phi)=0$ identically in our work.

\section{Hydrodynamic friction and active stresses}

 We now consider the effects of hydrodynamic friction, hitherto ignored, on 
the dynamics of a tensionless mixed membrane.
We include the effects of an active 
(nonequilibrium) stress $\sigma_{ij}=\gamma(\phi)p_i p_j$ (see, e.g., 
Ref.~\cite{sriram-RMP}), that is generically present in an active fluid. 
(This $\phi$-dependent active stress again reflects 
specific lipid dependence of the actin-lipid interactions.)
With $p_z=1,p_j=\partial_j 
h,j=x,y$, this makes a contribution of the form  $\partial_j 
(\gamma(\phi)p_zp_j)$ to $v_{hydro_z}$ at $z=h$; see Ref.~\cite{iso}. 
{We now choose $\gamma(\phi)=A\phi^2 + B\phi$.}
  {Because of their active origins, there 
are no 
restrictions on the magnitudes and signs of $A$ and $B$}. 
If hydrodynamic damping is considered, Eq.~(\ref{heq}) in 
the presence of the active stresses, modifies to [with the choice 
$X(\phi)=\Gamma_h (\lambda\phi^2+\tilde\lambda\phi)$, see above.]
%(set $\sigma =0$) {\bf NO 
%$\sigma$ in Eq. 3 there}
\begin{eqnarray}\label{extrah1}
 \frac{\partial h}{\partial t} &=& -\mu_p \frac{\delta {\mathcal 
F}}{\delta 
h} + \Gamma_h[\lambda\phi^2\nabla^2 h+ \tilde\lambda\phi\nabla^2 h] \nonumber\\ 
&+&
\Gamma_h^{\prime} [-\frac{\delta {\mathcal F}}{\delta 
h} + A{\boldsymbol\nabla}\cdot(\phi^2 {\boldsymbol\nabla}h)\nonumber \\ &+& B 
{\boldsymbol\nabla}\cdot(\phi {\boldsymbol\nabla}h)] 
+ f_h,
\end{eqnarray}
where $\Gamma_h^{\prime}$ is a damping coefficient. The terms within square 
brackets in (\ref{extrah1}) 
 with coefficients $\Gamma_h^{\prime}$ come from the 
solution of the 
three-dimensional hydrodynamic velocity field ${\bf v}_{hydro}$~\cite{iso}: 
\begin{equation}
 v_{hydro_z}=\Gamma_h^{\prime} [-\frac{\delta {\mathcal F}}{\delta 
h} + A{\boldsymbol\nabla}\cdot(\phi^2 {\boldsymbol\nabla}h) + B 
{\boldsymbol\nabla}\cdot(\phi {\boldsymbol\nabla}h)].\label{eqactivestress}
\end{equation}
For a constant $\Gamma_h^{\prime}$, active terms with coefficients $A$ and $B$ 
in (\ref{eqactivestress}) are total derivatives, and hence  
are subdominant (in a scaling sense) to those with coefficients 
$\lambda,\tilde\lambda$ in the hydrodynamic limit. Then, neglecting these $A$-
and $B-$ terms, we obtain
Eq.~(\ref{heq}) above.
For hydrodynamic friction, $\Gamma_h^{\prime}=1/(4\eta q)$ in the Fourier 
space, 
where $\eta$ is the 
ambient fluid viscosity (see 
Refs.~\cite{kremers,cai-lubensky}). As a result the $A$- and $B$-terms  no 
longer vanish in the hydrodynamic limit $q\rightarrow 0$, and hence compete 
with the active terms with coefficients $\lambda,\tilde\lambda$. Then, with
$\Gamma_h^{\prime}=1/(4\eta q)$,  
the $A$- and $B$-terms contain {\em fewer} derivatives than the 
$\lambda,\tilde\lambda$-terms and in a scaling sense should dominate over the 
$\lambda$- and $\tilde\lambda$-terms, respectively. This may 
be shown in a formal way. Replace $\phi^2$ by $\langle\phi^2\rangle$ and $\phi$ 
by $\langle\phi\rangle=0$ in the rhs of (\ref{extrah1}) above in a mean-field 
like approximation. This allows us 
to extract {\em two} active tensions:
\begin{equation}
 \sigma_a^h=A\langle\phi^2\rangle,\;\;\sigma_a=\lambda\langle\phi^2\rangle.
\end{equation}
{The former is the {\em active hydrodynamic tension}, where as the 
latter one is the active nonhydrodynamic tension [see Eq.~(\ref{activeten}). }
Equivalently, comparing with a tensionless isolated fluid membrane, we define 
an effective 
hydrodynamic
bending modulus $\kappa_e^h=\kappa +A\langle\phi^2\rangle/q^2$, in analogy 
with Eq.~(\ref{kappaeff}).
 In the long wavelength limit, in terms of 
ignoring the nonlinear terms, the effective dynamics of $h$ in the Fourier 
space is given by
\begin{equation}
 \frac{\partial h ({\bf q},t)}{\partial t} = -[\mu_p\kappa_e (q) 
q^4+\frac{\kappa_e^h}{4\eta} q^3] h ({\bf q},t) + f_h'. \label{eqh-hydro}
\end{equation}
The zero-mean, Gaussian white noise $f_h'$ should 
have a variance given by
\begin{equation}
 \langle |f_h'({\bf q},\omega)|^2\rangle = 2D_h + 2D_h'/q,
\end{equation}
$D_h'>0$.
Equation (\ref{eqh-hydro}) then yields for the 
equal-time membrane height correlator
\begin{equation}
 \langle |h({\bf q},t)|^2\rangle = \frac{D_h+D_h'/q}{\mu_p\kappa_e q^4 
+\kappa_e^h q^3/(4\eta)}.\label{hcorr-hydro}
\end{equation}
The active coefficients $\lambda$ and $A$ are formally 
independent of each other and can be positive or negative separately. As a 
result, a 
variety of situation may emerge. (i) If $\lambda$ and $A$ have the same sign - 
both positive or negative, we may neglect the $\lambda$-term in comparison with 
the $A$-term, and $\Gamma_h\kappa q^4$ in comparison with $\kappa q^3/(4\eta)$ 
in (\ref{hcorr-hydro}). Thus the long wavelength fluctuations of $h({\bf q},t)$ 
is now controlled by $\kappa_e^h$. Since
 $A$ can be positive or negative (just 
like $\lambda$), $\kappa_e^h$ varies with $A$ yielding
an $L$-dependence similar to that 
for $\kappa_e$ in the main text.  This evidently 
yields similar scaling
behaviours for $\Delta_n$ and $\Delta_h$ (as defined in the main text) across 
first order MPT 
and second order MPT, with now $A$ playing the role of $\lambda$ in the main 
text. Thus, 
the correspondence between MPTs and the membrane conformation fluctuations that 
is elucidated above with just nonhydrodynamic friction, survives with 
hydrodynamic friction as well. Furthermore, 
 for an impermeable membrane (for which $v_{perm}=0$, i.e., 
$\lambda=0=\tilde\lambda$) with 
hydrodynamic friction, $\Delta_n$ and $\Delta_h$ behave the same way as 
above across second order MPT and first order MPT, again establishing the 
direct correspondence 
between membrane conformation fluctuations 
and MPT, now for an impermeable membrane. (ii) Different signs of $A$ and 
$\lambda$: in this case, one would encounter instabilities. For instance, with 
$A<0$ and $\lambda>0$, the model displays instabilities at the smallest 
wavevectors, where for the opposite case ($A>0,\lambda <0$), the system remains 
stable at the 
smallest wavevectors, but shows finite wavevector instabilities controlled by 
the relative magnitudes of $A$ and $\lambda$. To what degree $A$ and $\lambda$ 
can be independently controlled and the biological significance of these 
results can be studied numerically by using properly constructed atomistic 
models.

\end{appendix}

 %\section{Supplemental Material}


\begin{thebibliography}{99}
\bibitem{inv} In-vitro model lipid bilayers are typically inversion-symmetric; 
see Ref.~\cite{veatch}.
\bibitem{noneqtemp} For the full nonequilibrium model, $r$ is a 
temperature-like tunable parameter in the model; $r$ may 
be tuned by controlling the ambient temperature. Obviously, $r$ assumes the 
usual significance of thermodynamic temperature in the equilibrium limit.
\bibitem{omit} {{ To recover the 
correct equilibrium limit, one needs to drop the $\lambda-$ and 
$\tilde\lambda$-terms in (\ref{heq})  and reinsert 
 the $\lambda_1-$ and $\lambda_2-$ terms in (\ref{heq}). The 
absence of the $\lambda_1-$ and $\lambda_2-$ terms in 
(\ref{heq}) does not affect our analysis of the active membrane fluctuations 
in the long wavelength limit.} }
\bibitem{alpha} { In general, $X(\phi)$ should have a $\phi$-independent 
constant 
piece $\alpha$ that yields an effective surface tension term in (\ref{heq}); 
see below. { We ignore this for simplicity.}}
\end{thebibliography}
\end{document}